\documentclass[preprint,aps,a4paper,floatfix,showkeys]{revtex4}
\usepackage{t1enc}
\usepackage{amsmath,amssymb,mathtools}
\usepackage{hhline,bbm,psfrag}
\usepackage{booktabs}
\usepackage{graphicx}

\newcommand\Source{ {S} }
\newcommand\StwoD{ {\Source}_{\rm 2D} } 
\newcommand\Sswirl{ {\Source}_{\rm swirl} } 
\newcommand\Smean{ \langle{\Sswirl}\rangle } 
\newcommand\Sfluc{ \tilde{\Source}_{\rm swirl} } 

\newcommand\Bound{ {b} }
\newcommand\Bmean{ \langle{\Bound}\rangle } 
\newcommand\Bfluc{ \tilde{\Bound} } 

\newcommand\ptwoD{ p_{\rm 2D} }
\newcommand\pswirl{ p_{\rm swirl} }

\newcommand\ptwoDmod{ p_{\rm 2D}^s }
\newcommand\StwoDmod{ {\Source}_{\rm 2D}^s } 

\newcommand{\atc}[1]{ \left.{#1}\right\vert_c }
\newcommand{\atone}[1]{ \left.{#1}\right\vert_{r=1} }
\newcommand{\atzero}[1]{ \left.{#1}\right\vert_{z=0} }

\newcommand\QtwoD{ Q_{\rm 2D} }
\newcommand\PtwoD{ P_{\rm 2D} }

\newcommand\Qswirl{ Q_{\rm swirl} }
\newcommand\Pswirl{ P_{\rm swirl} }

\newcommand{\eqsref}[1]{Eqs.~(\ref{#1})}

\begin{document}

\title{A fluid mechanic's analysis of the teacup singularity}

\author{Dwight Barkley}

\affiliation{Mathematics Institute, University of Warwick, 
  CV4 7AL Coventry, United Kingdom} 

\date{\today}

\begin{abstract}

The mechanism for singularity formation in an inviscid wall-bounded fluid flow
is investigated.  The incompressible Euler equations are numerically simulated
in a cylindrical container. The flow is axisymmetric with swirl.  The
simulations reproduce and corroborate aspects of prior studies reporting
strong evidence for a finite-time singularity.  The analysis here focuses on
the interplay between inertia and pressure, rather than on vorticity.
Linearity of the pressure Poisson equation is exploited to decompose the
pressure field into independent contributions arising from the meridional flow
and from the swirl, and enforcing incompressibility and enforcing flow
confinement.  The key pressure field driving the blowup of velocity gradients
is that confining the fluid within the cylinder walls.  A model is presented
based on a primitive-variables formulation of the Euler equations on the
cylinder wall, with closure coming from how pressure is determined from
velocity. The model captures key features in the mechanics of the blowup
scenario.

\end{abstract}


\maketitle

\section{Introduction}

In 1926 Einstein published a short paper explaining the meandering of
rivers~\cite{Einstein}. He famously began the paper by discussing the
secondary flow generated in a stirred teacup -- the flow now widely known to
be responsible for the collection of tea leaves at the centre of a stirred cup
of tea. In 2014, Luo and Hou presented detailed numerical evidence of a
finite-time singularity at the boundary of a rotating, incompressible,
inviscid flow~\cite{LH_PNAS,LH_MMS}.  The key to generating this singularity
is the teacup effect.  The present work is not aimed at proving the existence
of a singularity for this flow, nor is it aimed at generating more highly
resolved numerical evidence for the singularity than already exists.  Rather,
I assume that the flow simulated by Luo and Hou genuinely develops a
singularity in finite time. My goal is to understand, from a fluid-mechanics
perspective, why.

%%%%%%%%%%%%%%%%%%%%%%%%%%%%%%%%%%%%%%%%%%%%%%%%%%%%%%%%%%%%%%%%%%%%%%%%%%%%%

\section{Preliminaries}
\label{sec:preliminaries}

\subsection{Problem statement}

\begin{figure}%[tbhp]
\centering
\includegraphics[width=0.6\linewidth]{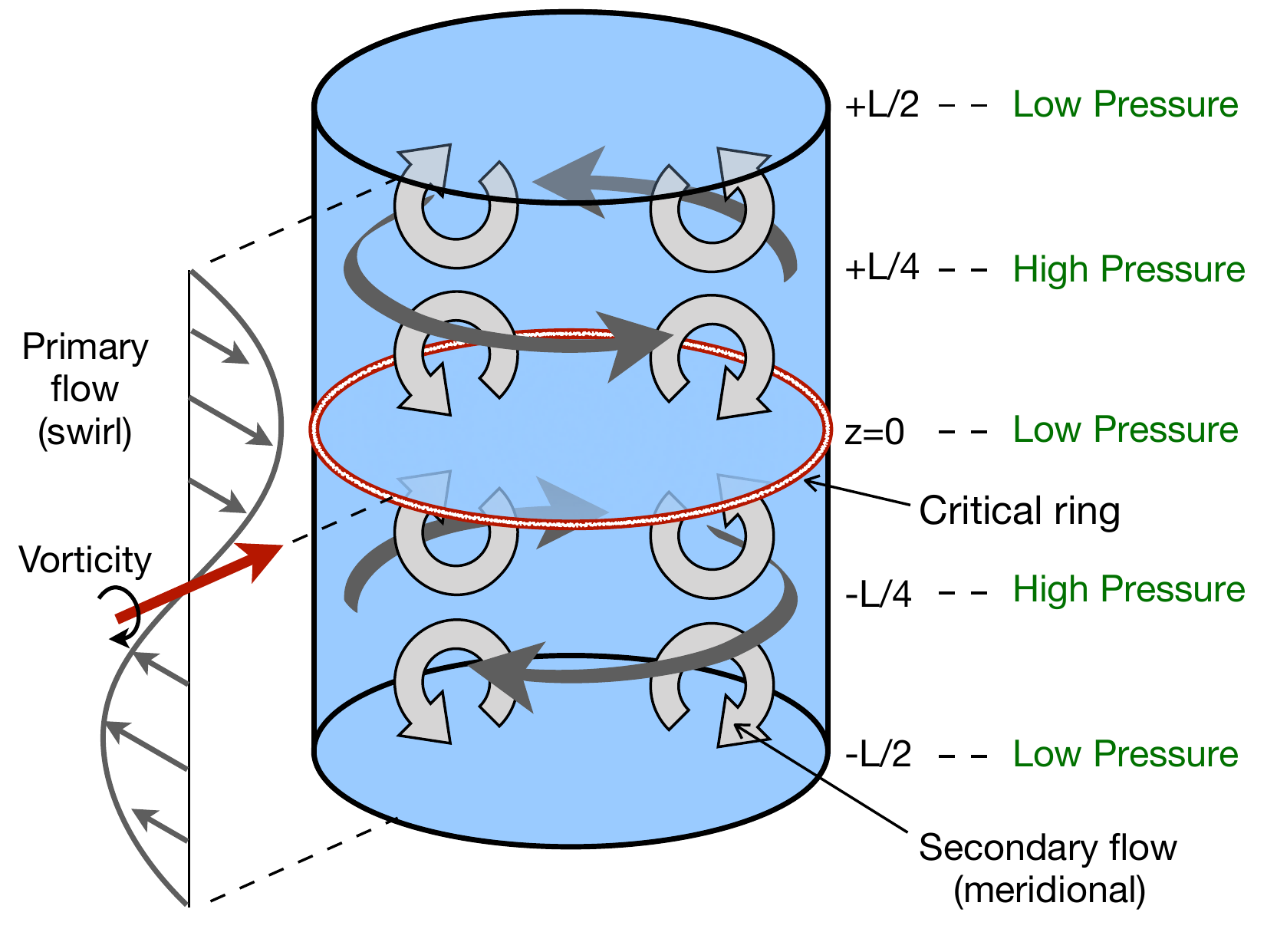}
\caption{Inviscid fluid flow in a cylinder periodic in the axial
  direction. The primary azimuthal flow (swirl) generates an axial
  variation in the pressure.  This produces a secondary meridional flow that
  in turn drives azimuthal flow along the cylinder wall towards the critical
  ring at $z=0$.  The shear of this azimuthal flow generates intense vorticity
  on the critical ring, ultimately leading to a singularity and a breakdown of
  the Euler equations.
  Note that by symmetry, a second critical ring (not indicated) exists at
  $z=L/2$, which by periodicity is also at $z=-L/2$. The portion of the fluid
  just from $z=0$ to $z=L/4$ corresponds to a cup of tea.  In the actual
  configuration studied, the height $L$ is only one sixth of the radius. See
  Fig.~\ref{fig:visu_large}. }
\label{fig:intro}
\end{figure}

The flow under investigation is depicted in Fig.~\ref{fig:intro}.  The system
is initialised with a pure azimuthal flow (swirl) having a sinusoidal
dependence on the axial coordinate $z$.  A pressure field is instantaneously
generated to provide the radially inward force necessary to keep fluid parcels
moving along circular paths.  This results in high pressure at the cylinder
wall where the circulation is largest ($z=\pm L/4$) and low pressure where
there is no azimuthal flow ($z=0$ and $z=\pm L/2$). Necessarily, then, there
is a vertical variation in the pressure at the cylinder wall and this drives a
secondary meridional flow. This is the teacup effect -- the portion of the
fluid just from $z=0$ to $z=L/4$ corresponds to a cup of tea.  (In an actual
cup of tea the variation in swirl with $z$ is due to a boundary layer at the
bottom of the cup.  Here we disregard viscous effects even though they play a
role in the flow of real tea in a teacup.)

%%%%%%%%%%%%%%%%%%%%%%%%%%%%%%%%%%%%%%%%%%%%%%%%%%%%%%%%%%%%%%%%%%%%%%%%%%%%%

%%%%%%%%%%%%%%%%%%%%%%%%%%%%%%%%%%%%%%%%%%%%%%%%%%%%%%%%%%%%%%%%%%%%%%%%%%%%%

We consider inviscid fluid flow governed by the incompressible Euler equations
\begin{subequations}
\begin{align}
  \partial_t u + u \cdot \nabla u & = -\nabla p, \label{eq:Euler_a} \\
  \quad \nabla \cdot u & = 0, \label{eq:Euler_b}
\end{align}
\label{eq:Euler}
% \end{subequations}
%
where $u$ is the fluid velocity and $p$ is the pressure divided by the fluid
mass density. By common usage we refer to $p$ simply as pressure.  We work in
cylindrical coordinates $(r,\theta,z)$.  The flow is axisymmetric (independent
of $\theta$), but has swirl ($u_\theta \ne 0$ in general). Hence the velocity
has components
$$
u(r,z,t) = u_r(r,z,t) \hat e_r + u_\theta(r,z,t) \hat e_\theta
+ u_z(r,z,t) \hat e_z,
$$
where $\hat e_r$, $\hat e_\theta$, and $\hat e_z$ are standard basis vectors
for cylindrical coordinates.  The vorticity is $\omega = \nabla \times u$ and
has corresponding components $\omega_r(r,z,t)$, $\omega_\theta(r,z,t)$, and
$\omega_z(r,z,t)$.  The flow takes place inside an axially periodic cylinder
of period $L = 1/6$ and radius 1.  The boundary condition at the cylinder wall
is
\begin{align}
\atone{u_r} = 0.
\label{eq:Euler_BC}
\end{align}
\end{subequations}

The initial condition employed by Luo and Hou, and reproduced here, is a
pure swirl
\begin{align}
u(r,z,t=0)
= 100 r e^{-30(1-r^2)^4} \sin\left(\frac{2\pi}{L}z\right) \hat e_\theta.
\label{eq:IC}
\end{align}
This initial condition possesses symmetries that are preserved under evolution
of \eqref{eq:Euler}. The most important is centro symmetry about $z=0$
$$
\left( u_r,  u_\theta, u_z\right) (r,z,t) 
=
\left( u_r, -u_\theta, -u_z\right) (r,-z,t).
$$
The full set of symmetry planes is $z_j = jL/4$, $j = 0, \pm 1, \pm 2$; $u_r$
is even and $u_z$ is odd about these all planes; $u_\theta$ is odd about
planes $z_o$, $z_{\pm 2}$ and is even about planes $z_{\pm 1}$. The pressure
$p$ is even about all four planes.

Extensive analysis of finely resolved numerical simulations of the Euler
equations indicates that starting from the above initial condition, the flow
evolves to form a singularity on the critical ring, $(r=1, z=0)$, at time $T
\simeq 0.0035056$ ~\cite{LH_PNAS,LH_MMS,LH_Review}.  In the present work,
simulations are well resolved to time $t = 0.0031$.  Details of the
simulations are given in Appendix D.  I rely heavily on the studies of Luo and
Hou (hereafter referred to as LH), to know that the flow at $t = 0.0031$ is
indicative of the flow all the way to $t = 0.003505$, extremely close to the
singularity time.  To be clear, the simulations presented here are not aimed
at numerically establishing a singularity (LH have already done this), but
instead at understanding the physical mechanisms at work, and for this purpose
they are adequate.

\subsection{Mechanics}
\label{sec:mechanics}

Pressure is the only stress acting within an inviscid fluid and it is the only
means to provide force to, and thereby accelerate, the flow.  It is at the
heart of the teacup effect and it is therefore natural to investigate its role
in the singularity.  In general, stress is a tensor field $\tau$ whose
divergence gives the net force acting on infinitesimal fluid parcels.
Pressure is isotropic and so for inviscid flow $\tau$ has only diagonal
components: $\tau_{ij} = -P \delta_{ij}$, where $P$ is the pressure field, and
$\delta_{ij}$ is the Kronecker delta.  Hence $\nabla \cdot \tau = -\nabla P$
is the force per volume acting within the fluid.  For incompressible flow, the
fluid's mass density $\rho$ is constant and we define $p \equiv P/\rho$, so
that $-\nabla p$ sets the fluid acceleration and thus appears on the
right-hand-side of the momentum equation \eqref{eq:Euler_a}.  Subsequently the
symbol $P$ will be used for another quantity and we will refer to $p$ simply
as pressure.  Pressure gradients with $-\nabla p$ anti-parallel to velocity
$u$ are known as adverse pressure gradients and result in flow deceleration
(decrease in fluid speed).

The role of pressure in incompressible flow is seen by taking the divergence
of \eqref{eq:Euler_a}
\begin{equation}
  \partial_t \left( \nabla \cdot u \right) + \nabla \cdot \left( u \cdot
  \nabla u \right) = -\nabla^2 p. 
\label{eq:dilatation}
\end{equation}
This equation governs the evolution of the flow divergence.  Given a
divergence-free velocity field $u$ satisfying \eqref{eq:Euler_b}, in general
$\nabla \cdot \left( u \cdot \nabla u \right)$ will not be zero, meaning that
nonlinearity acting alone does not maintain incompressibility.  A pressure
field is generated within the fluid (simultaneously everywhere) to accelerate
the flow exactly so as to counterbalance this effect of nonlinearity.  From
\eqref{eq:dilatation}, the relationship between pressure and velocity required
to maintain a divergence-free flow is the Poisson equation
\begin{equation*}
\nabla^2 p = -\nabla \cdot \left( u \cdot \nabla u \right).
\end{equation*}

This is not the full story, however. The flow of interest is wall bounded and
this puts a condition on the stress field within the fluid.  The initial
velocity field satisfies \eqref{eq:Euler_BC} and thus has no radial component
at the cylinder wall.  From the $\hat e_r$ component of the momentum equation
at the wall, this will be maintained as long as $\atone{\partial_r p} =
\atone{u_\theta^2}$.  Thus, pressure is determined by a Poisson equation
together with its boundary condition
\begin{equation}
  \nabla^2 p = -\nabla \cdot \left( u \cdot \nabla u \right)
  \equiv \Source, \quad 
\atone{\partial_r p} = \atone{u_\theta^2} \equiv \Bound,
\label{eq:PPE}
\end{equation}
%
% See Preston and Sarria (2016) and Pope 
%
where these expressions define the source term $\Source$ and the boundary term
$\Bound$.  As long as $p$ satisfies \eqref{eq:PPE}, the flow evolving under
\eqref{eq:Euler_a} will remain incompressible and confined within the
cylinder.  An important focus of this work will be disentangling the
contributions to the stress associated with incompressibility from those
associated with flow confinement.

%%%%%%%%%%%%%%%%%%%%%%%%%%%%%%%%%%%%%%%%%%%%%%%%%%%%%%%%%%%%%%%%%%%%%%%%%%%%%

\section{Basics of the singularity mechanism}

\begin{figure}
\centering
\includegraphics[width=1.0\linewidth]{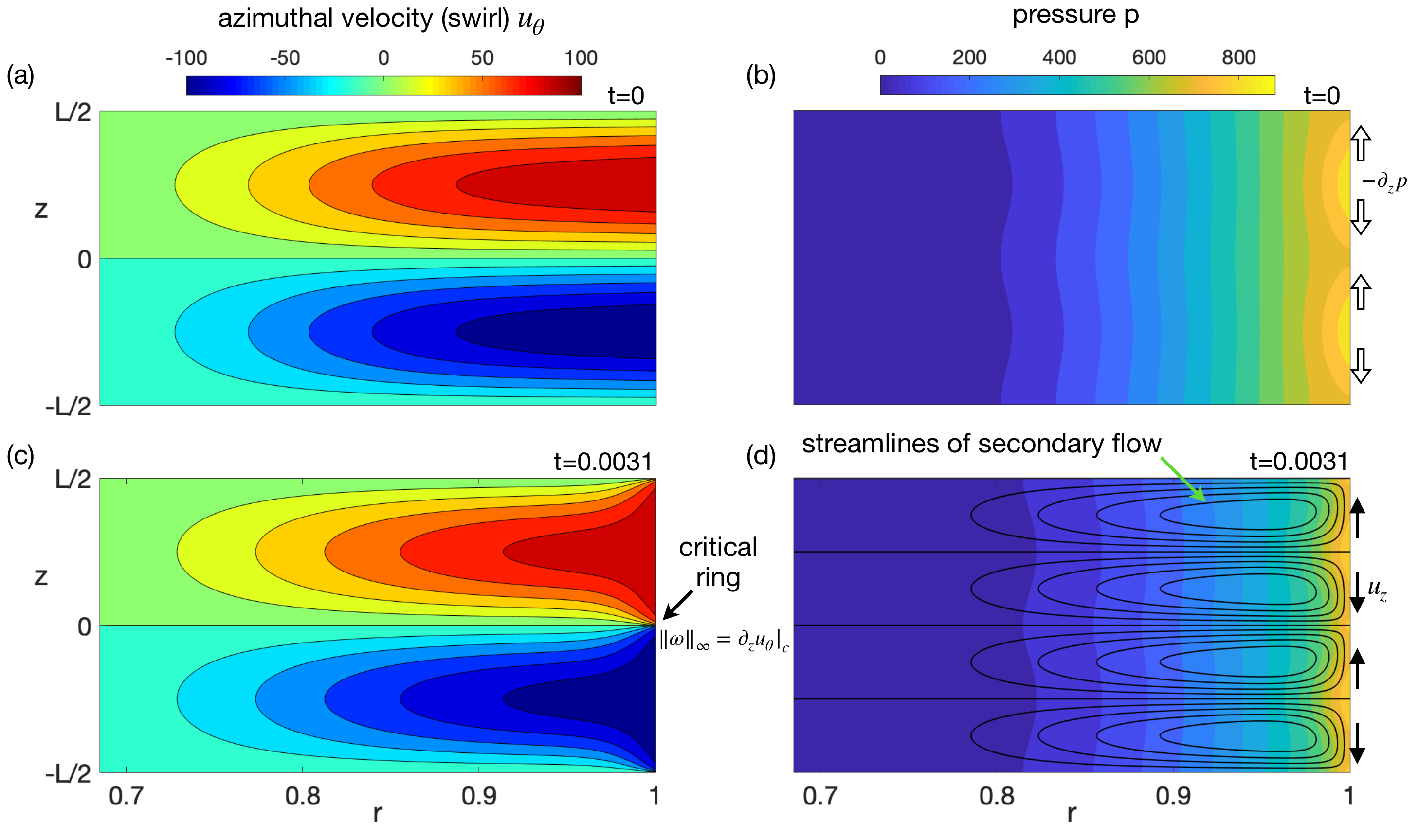}
\caption{Overview of the flow dynamics -- similar to the sketch in
  \ref{fig:intro} but for the actual flow configuration.
  (a, b) Initial azimuthal velocity (swirl) from \eqref{eq:IC} and
  corresponding pressure field in a meridional plane.  The initial swirl is
  concentrated near the cylinder wall $(r=1)$ and only the outer third of the
  radius is shown. Axial variation of this primary flow results in an
  axial pressure gradient (indicated by open arrows) that drives a secondary
  meridional flow.
  (c, d) Azimuthal flow, pressure field, and secondary meridional flow at the
  final simulation time $t=0.0031$, the standard case analysed in this paper.
  The meridional flow is shown by contours of the Stokes streamfunction.  The
  surfaces $z=0$, $z=\pm L/4$, $z=\pm L/2$, and $r=1$ are flow invariant.
  Arrows indicate the direction of the meridional flow along the cylinder
  wall.  Advection of $u_\theta$ along the cylinder wall by the secondary flow
  results in intense radial vorticity $\omega_r = -\partial_z u_\theta$ on the
  critical ring $(r=1,z=0)$.
}
\label{fig:visu_large}
\end{figure}

Figure~\ref{fig:visu_large} presents a quantitative overview of the flow
dynamics with visualisation of the initial and final states from the numerical
simulation.  We see the teacup effect: the initial primary flow (pure swirl)
results in high pressure on the cylinder wall at $z=\pm L/4$, where the swirl
is largest. This in turn results in a vertical component of the pressure
gradient that produces a secondary flow driving fluid on the cylinder wall
toward the midplane $z=0$, (and by symmetry also toward $z=\pm L/2$).  Swirl
of opposite signs from above and below the midplane is thus transported
towards $z=0$, resulting in intense radial vorticity $\omega_r = -\partial_z
u_\theta$ on the critical ring.  This has been described clearly by
LH~\cite{LH_PNAS,LH_MMS,LH_Review}, who show very strong evidence that the
flow continues to develop a singularity in a nearly, (but not exactly
~\cite{chae_Tsai_2015,Sperone_2017}), self-similar way.  The final state of
the present simulations shown in Fig~\ref{fig:visu_large} is representative of
the flow as it approaches the singularity.

%%%%%%%%%%%%%%%%%%%%%%%%%%%%%%%%%%%%%%%%%%%%%%%%%%%%%%%%%%%%%%%%%%%%%%%%%%%%%
%
% for reference:
% max omega_theta = 2.103538590000000e+03
% max omega_z = 1.787519170000000e+04
%
%%%%%%%%%%%%%%%%%%%%%%%%%%%%%%%%%%%%%%%%%%%%%%%%%%%%%%%%%%%%%%%%%%%%%%%%%%%%%

\begin{figure}
\centering
\includegraphics[width=1.0\linewidth]{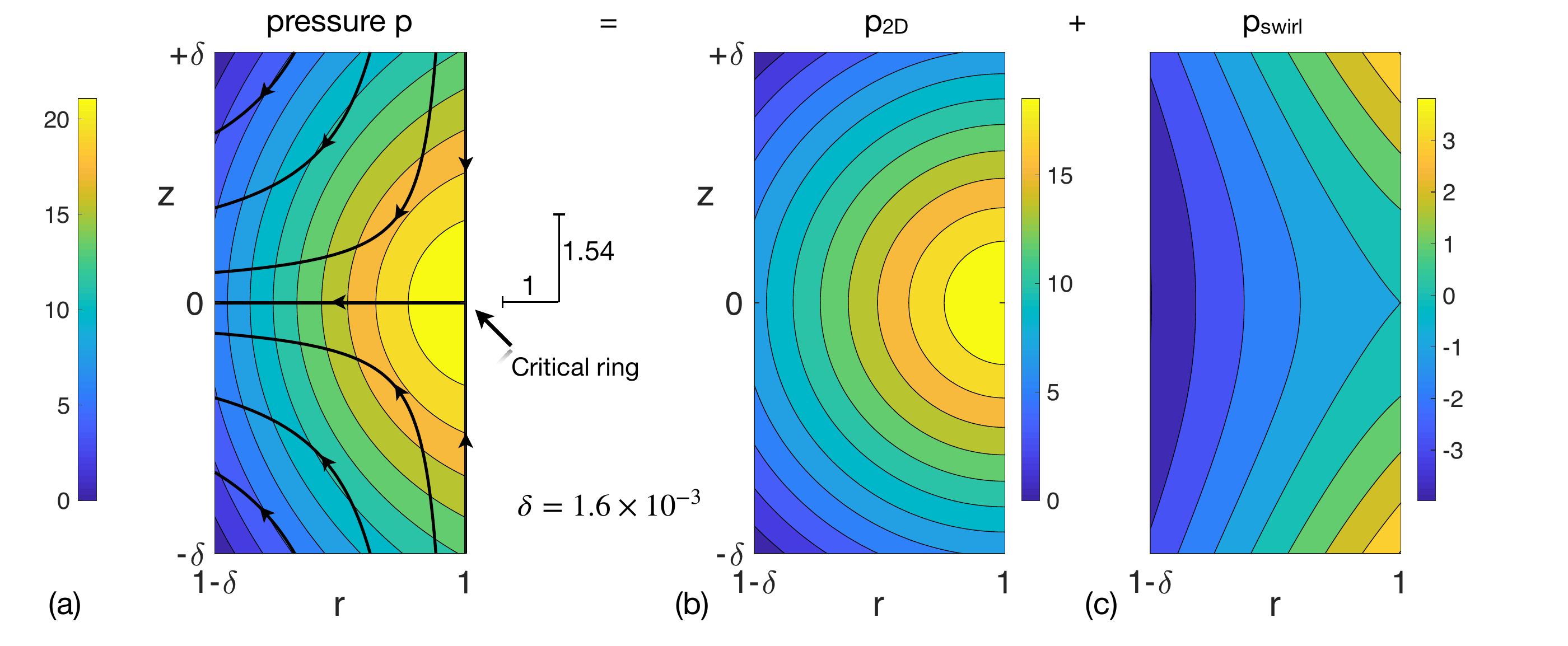}
  \caption{(a) Enlargement very near the critical ring of the pressure field
    and meridional-flow streamlines from Fig.~\ref{fig:visu_large}(d). (This
    region, with $\delta = 1.6\times 10^{-3}$, will be used in all subsequent
    plots.)  The critical ring is a saddle point for the meridional flow and a
    local pressure maximum diverts (accelerates) the incoming flow. In
    particular, the pressure decelerates axial flow $u_z$ on the cylinder wall
    as it converges towards the critical ring.  The length ratio 1.54-to-1
    associated with exponent $\gamma = 2.46$ is indicated (see text).
    (b) Pressure $p$ is decomposed into the sum of $\ptwoD$ and $\pswirl$,
    where $\ptwoD$ is determined from the meridional (2D) flow and $\pswirl$
    from the swirl $u_\theta$.  The contours of $\ptwoD$ are nearly circular
    arcs centred on the critical ring, while $\pswirl$ is a hyperbolic point
    (saddle) with high pressure along the cylinder wall.
}
\label{fig:p_p2D_pswirl}
\end{figure}

Figure \ref{fig:p_p2D_pswirl}(a) shows an enlargement of the final flow in a
very small region around the critical ring.  Analysis of this state will be a
major focus of the paper.  The critical ring is a saddle point of the
meridional flow.  A local pressure maximum exists on the critical ring (barely
visible in Fig.~\ref{fig:visu_large}). This secondary local maximum in the
pressure field first appears at time $t \simeq 0.002$ and accounts for the
stress required to accelerate the flow around the saddle point -- bending
incoming axial velocity near the cylinder wall to radially inward velocity
near the midplane.  This pressure field is similar to that reported by LH at
$t=0.003505$, very close to the singularity time $T \simeq 0.0035056$. (See
Ref.~\cite{LH_MMS}, but note that its Fig.~17 has a distorted aspect ratio.)
LH emphasise that the pressure maximum on the critical ring means that there
is locally an adverse axial pressure gradient decelerating the incoming axial
flow on the cylinder wall.  This is an important point. However, it does not
mean that the pressure maximum inhibits the singularity. On the contrary, a
pressure maximum like that in Fig.~\ref{fig:p_p2D_pswirl}(a) can drive a
singularity.  This fact is central to this work.

%%%%%%%%%%%%%%%%%%%%%%%%%%%%%%%%%%%%%%%%%%%%%%%%%%%%%%%%%%%%%%%%%%%%%%%%%%%%%

%%%%%%%%%%%%%%%%%%%%%%%%%%%%%%%%%%%%%%%%%%%%%%%%%%%%%%%%%%%%%%%%%%%%%%%%%%%%%

%%%%%%%%%%%%%%%%%%%%%%%%%%%%%%%%%%%%%%%%%%%%%%%%%%%%%%%%%%%%%%%%%%%%%%%%%%%%%

To understand the mechanics of this particular situation, we turn to the
velocity-gradient dynamics on the critical ring.  Differentiating velocity
gives the velocity-gradient tensor $\nabla u$ and differentiating the pressure
gradient gives the pressure Hessian $\nabla (\nabla p)$.  Symmetries dictate
that on the critical ring the only non-zero derivatives entering these are
\begin{subequations}
\begin{equation}
W \equiv \atc{\partial_z u_z}, \quad 
\Omega \equiv \atc{\partial_z u_\theta}, \quad
V \equiv \atc{\partial_r u_r}, 
\end{equation}
\begin{equation}
P \equiv \atc{\partial_{zz} p}, \quad  
Q \equiv \atc{\partial_{rr} p}, 
\end{equation}
\end{subequations}
where $\atc{}$ means evaluated on the critical ring.  $P$ and $Q$ will be
especially important in what follows. I refer to these as {\em pressure
  curvatures} since they are the principal curvatures of a graph of the
pressure $p(r,z)$. It is to be understood that I always mean ``on the critical
ring'' when referring to these curvatures.  Straightforward differentiation of
\eqref{eq:Euler_a} gives
\begin{equation*}
\dot W + W^2 = -P,  \quad 
\dot \Omega + W \Omega = 0, \quad 
\dot V + V^2 = -Q. 
\end{equation*}
By incompressibility on the critical ring: $V + W = 0$.  Thus, $V$ can be
eliminated, giving the velocity-gradient dynamics
\begin{subequations}
\label{eq:main}
\begin{alignat}{3}
&  \dot W + W^2 = -P,
&\quad  & \mbox{(axial momentum)}
  \label{eq:Wdot} \\
& \dot \Omega + W \Omega = 0,
& & \mbox{(vortex stretching)}
  \label{eq:Omegadot} \\
&  P + Q = -2 W^2
& & \mbox{(pressure Poisson or mean curvature)}
  \label{eq:PPEc} 
\end{alignat}
\end{subequations}
The meaning associated with each equation is indicated.  The equations are
exact, and while they are not closed (\eqref{eq:PPEc} is insufficient to
determine $P$ and $Q$ separately), they are extremely useful in examining what
transpires in singularity formation.  For this flow, $-\atc{\omega_r} =
\atc{\partial_z u_\theta} = \Omega$ is the absolute vorticity
maximum~\cite{LH_PNAS,LH_MMS}, so $\Omega = \| \omega \|_\infty =
\atc{\partial_z u_\theta} $ as indicated in Fig.~\ref{fig:visu_large}.
Equation \eqref{eq:PPEc} is the pressure Poisson equation evaluated on the
critical ring, but equally it is the geometrical statement that sum of
principal curvatures is twice the mean curvature, where the mean curvature of
$p$ is $-W^2$.
(See works by D.\ Chae and collaborators
\cite{chae2008incompressible,chae2008blow,constantin2008singular,
  chae2010lagrangian, Chae_etal_2012} for more general treatments of the Euler
equations in the velocity-gradient formulation, including several 
blowup scenarios.)

From Fig.~\ref{fig:p_p2D_pswirl}(a) we see that the principal pressure
curvatures, $P$ and $Q$, are both negative (a pressure maximum occurs on the
critical ring), but that they are not equal.  The axial curvature is smaller
in magnitude than the radial curvature, that is $|P| < |Q|$.  This can be seen
in the ratio of axial to radial length scales in the pressure contours. This
leads us to define $a \ge 0$ by
\begin{equation}
a^2 \equiv Q/P,
\label{eq:a2} 
\end{equation}
so that $a$ is this ratio of length scales.  The case $a > 1$ corresponds to
$|P| < |Q|$ and is seen in Fig.~\ref{fig:p_p2D_pswirl}(a).

To understand the importance of $|P| < |Q|$ to blowup, we proceed as follows.
Using \eqref{eq:a2} to eliminate $Q$ from \eqref{eq:PPEc} gives $P = -2
W^2/(a^2+1)$, which can then be used to eliminate $P$ from
\eqref{eq:Wdot}. The velocity-gradient equations \eqref{eq:main} then become
\begin{equation}
\dot W = -\frac{W^2}{\gamma}, \quad \dot \Omega = -W
\Omega \label{eq:main2}
\end{equation}
where
\begin{equation}
\gamma \equiv \frac{a^2+1}{a^2-1} = \frac{Q+P}{Q-P}
\label{eq:gamma}
\end{equation}
The case of interest $\infty > a > 1$ corresponds to $1 < \gamma < \infty$.

The solution to \eqsref{eq:main2} with $\gamma$ constant is simple and
captures the essence of the blowup described by these equations.  Effectively
we set $\gamma$ to its limiting value, assumed to be finite, as $t \to T$.  We
are interested in a saddle point flow in the meridional plane, with fluid
converging towards the critical ring in the axial direction. We are only ever
interested in this situation and always assume $W(t=0) = W_0 < 0$.  The
solution to \eqsref{eq:main2} is then
\begin{equation}
  W(t) = -\frac{\gamma}{T-t} \sim (T-t)^{-1}, 
  \quad \Omega(t) = 
  \frac{\Omega_0 T^\gamma}{(T-t)^\gamma} \sim (T-t)^{-\gamma}, 
  \label{eq:sol}
\end{equation}
where $T = -\gamma/W_0 > 0$ is the singularity time and $\Omega_0 =
\Omega(0)$.  These are the known divergences as $t \to
T$~\cite{LH_PNAS,LH_MMS}.  In particular, the vorticity $\Omega = \|\omega
\|_\infty$ diverges with exponent $-\gamma$.  All other divergences associated
with the singularity follow from invariances of the Euler equations and the
value of $\gamma$.  By treating $\gamma$ as a constant, we obtain the scaling
of the blowup as a simple exact solution to \eqref{eq:main2}. This will be
useful in what follows.

We know from LH that the vorticity diverges with exponent $\gamma \simeq
2.46$, corresponding to $a \simeq 1.54$. The corresponding ratio of length
scales is indicated in Fig.~\ref{fig:p_p2D_pswirl}(a).  The contours do not
exactly manifest this ratio of scales, in part because contours are a finite
distance from the critical ring and in part because the flow is seen at a time
a finite distance from the singularity time.  From the data at $t=0.0031$,
$\sqrt{Q/P} \simeq 1.62$.  (See data in Table \ref{tab:curvatures}.)
%

% From Matlab with file Pfull:
% Q = DDPr_0 = -5.187199464875035e+07
% P = DDPz_0 = -1.987654162187328e+07
% a^2 = 2.609709256044193e+00
% a =   1.615459456638944e+00
% gamma = 2.242460396180447e+00

The fundamental point is the following. Incompressibility locks axial
contraction and radial expansion together such that it is not the signs of
principal pressure curvatures $P$ and $Q$ that are important for singularity
formation; it is their inequality. A persistent inequality in the pressure
curvatures on the critical ring can drive the flow to a singularity.  Of
interest here is flow converging axially toward the critical ring with $|P| <
|Q|$, so the axial curvature is smaller than the radial curvature in
magnitude.  The pressure contours in Fig.~\ref{fig:p_p2D_pswirl}(a) are the
signature of this simple mechanism. If the flow evolves such that this
situation persists, (that is such that $\inf_{t \ge 0} Q/P > 1$), then the
solution will blow up.  One can deduce from the results of LH that a ratio of
pressure curvatures of approximately the same amount as is seen in
Fig.~\ref{fig:p_p2D_pswirl}(a) exists as close to the singularity time as they
could resolve (Fig.~17 of Ref.~\cite{LH_MMS}).

%%%%%%%%%%%%%%%%%%%%%%%%%%%%%%%%%%%%%%%%%%%%%%%%%%%%%%%%%%%%%%%%%%%%%%%%%%%%%

\section{Illustrative cases}
\label{sec:illustrative}

Before continuing to a detailed analysis of the pressure field, I consider the
the velocity-gradient dynamics \eqref{eq:main} in two limiting cases.  These
cases will appear again later in the paper (see
Fig.~\ref{fig:u_theta_Omega_W}), and they are very useful for understanding
the interplay between inertia and pressure in the mechanics of the
singularity.

Consider simply dropping pressure and the incompressibility constraint from
the Euler equations, and also dropping the radial component for the momentum
equation. On the cylinder wall, the remaining two components of the momentum
equation become Burger's equation and advection of swirl as a passive scalar:
$$
\partial_t u_z + u_z \partial_z u_z = 0, \quad
\partial_t u_\theta + u_z \partial_z u_\theta = 0.
$$
The velocity-gradient dynamics on the critical ring become
\begin{equation}
  \dot W + W^2 = 0, \quad \dot \Omega + W \Omega = 0.
\end{equation}
(Pressure does not appear and \eqref{eq:main}(c) is dropped.)  Starting with a
flow converging towards $z=0$, $W_0 < 0$, these equation have blowup given by
\begin{equation}
  W(t) = -\frac{1}{T-t}, \quad \Omega(t) = \frac{\Omega_0 T}{T-t},
  \label{eq:Burger_sol}
\end{equation}
where $T = -1/W_0> 0$ is the singularity time.  This is just a special case of
Eq.~\eqref{eq:sol} with $\gamma=1$.  In the absence of stresses, there is no
deceleration of the fluid parcels converging towards $z=0$, resulting in the
well-known blowup of Burger's equation. This illustrates how inertia, or
equivalently the associated advective nonlinearity in Eulerian coordinates,
itself can easily lead to a finite-time singularity.

Consider now the case in which principal pressure curvatures are equal at all
times: $P=Q$.  This would correspond to pressure contours locally forming
circular arcs about the critical ring in the meridional plane (similar to what
is seen in Fig.~\ref{fig:p_p2D_pswirl}(b), although those contours are not
perfectly circular).  With this assumption, Eqs.~\eqref{eq:main} are closed
because both pressure curvatures equal the mean curvature: $P = Q =
-W^2$. With this, the velocity-gradient dynamics on the critical ring become
\begin{equation}
  \dot W + W^2 = W^2, \quad \dot \Omega + W \Omega = 0.
  \label{eq:main2D}
\end{equation}
(This case corresponds to $\gamma=\infty$.)  The system does not develop a
singularity and instead has solution
\begin{equation}
W(t) = W_0, \quad \Omega(t) = \Omega_0 \exp(-W_0 t).
  \label{eq:2D_sol}
\end{equation}
Since $W_0 < 0$, the vorticity grows exponentially, but only exponentially in
time. This illustrates what is observed to be the normal situation for
incompressible inviscid flow -- the stress that develops within the flow to
maintain incompressibility is such as to accelerate the fluid sufficiently to
prevent blowup that would come from inertia acting alone.  Algebraically, the
term $W^2$ from inertia on the left-hand side of \eqref{eq:main2D} is exactly
balanced by the term $W^2$ from pressure on the right-hand side.

The case of interest, where $|P| < |Q|$ and hence $1 < \gamma < \infty$, falls
between the two extremes just considered.  We have a flow configuration
evolving under the full incompressible Euler equations, with the pressure
stress acting, but such that the axial pressure curvature $P$ is too small to
compensate inertia. As a result, fluid parcels converging towards $z=0$ are
not sufficiently decelerated and a singularity occurs.

%%%%%%%%%%%%%%%%%%%%%%%%%%%%%%%%%%%%%%%%%%%%%%%%%%%%%%%%%%%%%%%%%%%%%%%%%%%%%

\section{Analysis of pressure}
\label{sec:pressure}

In this section I will analyse in depth the structure of the pressure field
near the critical ring and show how it is dictated by specific aspects of the
fluid flow.  I will then use this information in Sec.~\ref{sec:model} to gain
further insights into the blowup scenario.

\subsection{Meridional and swirl pressure fields}

We exploit the linearity of the Poisson equation \eqref{eq:PPE} to separate
pressure into contributions from distinct effects.  To begin, the source term
for the equation can be decomposed as $\Source = \StwoD + \Sswirl$, where
$\StwoD$ depends only on the meridional (2D) velocity components $(u_r, u_z)$
and $\Sswirl$ depends only on the swirl velocity $u_\theta$. (See Appendix A
for details.) The boundary term $\Bound$ in \eqref{eq:PPE} also depends only
on $u_\theta$.  Thus, the pressure $p$ can be written as a linear
superposition $p = \ptwoD + \pswirl$, where
\begin{subequations}
\begin{alignat}{3}
\nabla^2 \ptwoD &= \StwoD, \quad &
 \left. \partial_r \ptwoD \right\vert_{r=1} & = 0, \label{eq:PPE2D} \\
\nabla^2 \pswirl &= \Sswirl, \quad & 
\left. \partial_r \pswirl \right\vert_{r=1} & = \Bound.
\label{eq:PPEswirl}
\end{alignat}
\label{eq:p2D_pswirl}
\end{subequations}

These pressure fields are plotted in Fig.~\ref{fig:p_p2D_pswirl}(b).  Contours
of $\ptwoD$ are nearly circular arcs indicating approximate rotational
symmetry locally about the critical ring within the meridional plane.
Contours of $\pswirl$ are those of a hyperbolic saddle with the expected high
pressure along the cylinder wall where the swirl is largest.  Since we will be
especially concerned with the axial momentum balance, these fields are plotted
along the cylinder wall in Fig.~\ref{fig:lines1}(a).

Let
$$
P = \PtwoD + \Pswirl, \quad Q = \QtwoD + \Qswirl, 
$$
where $\PtwoD = \atc{\partial^2_z \ptwoD}$, $\QtwoD = \atc{\partial^2_r
  \ptwoD}$, etc, are the principal curvatures of the component pressure
fields. (See Table \ref{tab:curvatures}.)  Since $\atc{\Sswirl}=0$, we have
from \eqref{eq:PPEswirl} and \eqref{eq:PPEc}
\begin{equation}
\PtwoD + \QtwoD = -2 W^2, \quad \Pswirl + \Qswirl = 0.
\label{eq:PQ_decomp}
\end{equation}
Hence the mean curvature of the pressure field $p$ is contained entirely in
the component field $\ptwoD$. (This is obvious since both $\ptwoD$ and the
mean curvature $-W^2$ are functions only of the meridional flow, and $\pswirl$
is not.)  Necessarily then the swirl pressure always has zero mean curvature.

The core cause for the inequality in the pressure curvatures, $|P| < |Q|$, is
immediately evident.  The near symmetry of the meridional pressure maximum
implies that $\PtwoD \simeq \QtwoD \simeq -W^2 < 0$, while for the saddle
swirl pressure $\Pswirl > 0 > \Qswirl$. Hence
\begin{equation}
|P| = |\PtwoD + \Pswirl| < |\QtwoD + \Qswirl| = |Q|. \label{eq:mismatch}
\end{equation}
Stated simply -- the pressure maximum from the meridional flow is flattened by
the swirl pressure the axial direction, but it is steepened by the swirl
pressure in the radial direction. This is seen in the visualisations of
Fig.~\ref{fig:p_p2D_pswirl} and shown quantitatively along the cylinder wall
in Fig.~\ref{fig:lines1}(a).  To exploit fully this insight, more detailed
information is required on the meridional and swirl pressure fields.

\begin{table}
\begin{tabular}{llcllcll}
\toprule
Curvature & ~~~~~~value & & Curvature & ~~~~~~value & &
Quantity & ~~~~~~value \\
$P$  & $-1.9877 \times 10^7$ & &
$Q$ & $-5.1872 \times 10^7$ & &
$\atc{\partial_r p_a}$ & $-8.6991 \times 10^3$ 
\\ \midrule
$\PtwoD$ & $-2.8556  \times 10^7$ & &
$\QtwoD$ & $-4.3192  \times 10^7$ & &  
$\atc{\partial_r p_c}$ & $ ~~~8.6991 \times 10^3$ 
\\  \midrule
$\Pswirl$ & $ ~~~8.6797 \times 10^6$ & &
$\Qswirl$ & $-8.6797 \times 10^6$ & &
$V$ & $  ~~~5.9895 \times 10^3$ 
\\  \midrule
$P_a$ & $ ~~~9.5265 \times 10^6$ & &
$Q_a$ & $-9.5178 \times 10^6$ & &
$W$ & $ -5.9895 \times 10^3$ 
\\ \midrule
$P_b$ & $ -8.4676 \times 10^5$ & &
$Q_b$ & $ ~~~5.7998 \times 10^5$ & &
$\Omega$ & $ ~~~1.5428 \times 10^5$ 
\\  \midrule
$P_c$ & $ ~~~0$ & &
$Q_c$ & $ ~~~2.5808 \times 10^5$ & & & \\  \midrule
\end{tabular}
\caption{Tabulated principal pressure curvatures and other quantities at
  $t=0.0031$.}
\label{tab:curvatures}
\end{table}

%%%%%%%%%%%%%%%%%%  2D text %%%%%%%%%%%%%%%%%% 

\subsection{Meridional pressure}
\label{sec:meridional}

The pressure field $\ptwoD$ exists within the fluid to counter divergences
that would be otherwise generated just by meridional flow. As the radial
gradient of this field is zero at the cylinder wall, \eqref{eq:PPE2D}, it does
not contribute to fluid confinement.  It is determined only by the
instantaneous state of the meridional flow.  In a region around the critical
ring the meridional velocity field is a saddle that is approximately
anti-symmetric under interchange of the axial and radial directions.  See the
streamlines in Fig.~\ref{fig:p_p2D_pswirl}(a) where the Stokes streamfunction
locally satisfies $\psi(r,z) \simeq \psi(1-z,1-r)$ for $z \ge 0$, and
similarly for $z\le 0$.  Although the flow cannot globally respect such a
symmetry, very near to the critical ring it does, approximately.  Such a
saddle flow is to be anticipated
\cite{Kiselev_Sverak_2014,CKY_2015,Kiselev_Tan_2018} and the associated
approximate rotational symmetry of $\ptwoD$ near a local maximum is not
particularly surprising. The source term $\StwoD$ is quadratic in velocity so
an approximately anti-symmetric streamfunction implies an approximately
symmetric pressure field.

\begin{figure}[tbp]
\centering
\includegraphics[width=1.0\linewidth]{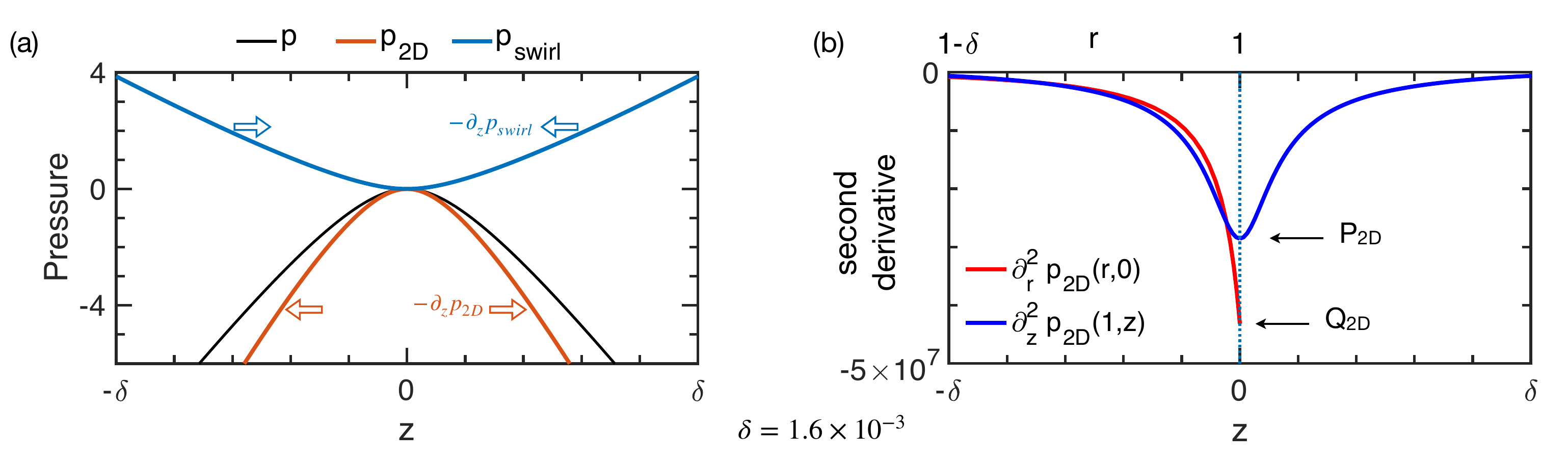}
\caption{
  (a) Pressure components from Fig.~\ref{fig:p_p2D_pswirl} plotted as function
  of $z$ along the cylinder wall $(r=1)$.  Arbitrary constants are such that
  fields are all zero at $z=0$.  The meridional pressure $\ptwoD$ has negative
  curvature (local maximum) and hence $-\partial_z \ptwoD$ is an adverse
  pressure gradient directed outward from the critical ring (open red arrows).
  The axial curvature of the swirl pressure $\pswirl$ is positive and hence
  $-\partial_z \pswirl$ is a favourable pressure gradient directed toward the
  critical ring (open blue arrows).  The full pressure field $p =
  \ptwoD+\pswirl$ has a local maximum, but with less axial curvature than
  $\ptwoD$, and hence a weaker adverse pressure gradient (not indicated), than
  $\ptwoD$ near the critical ring.
  (b) Second derivatives of $\ptwoD$ along one-dimensional slices:
  $\partial^2_r\ptwoD$ on the midplane (red) and $\partial^2_z\ptwoD$ on the
  cylinder wall (blue).  The dotted line indicates the critical ring in both
  cases, where the second derivatives give the pressure curvatures $\PtwoD$
  and $\QtwoD$.  (Numerical values for the curvatures are given in Table
  \ref{tab:curvatures}.)  The near symmetry of $\ptwoD$ in the radial and
  axial directions does not hold on the critical ring where $|\PtwoD| <
  |\QtwoD|$.
}
\label{fig:lines1}
\end{figure}

However, it is the principal pressure curvatures on the critical ring that
matter for singularity formation. So while the near symmetry of $\ptwoD$ seen
in Fig.~\ref{fig:p_p2D_pswirl}(b) suggests that $\PtwoD \simeq \QtwoD$, it is
necessary to examine these curvatures quantitatively, in particular to
understand in what way they are not exactly equal.  Figure~\ref{fig:lines1}(b)
shows second derivatives of $\ptwoD$ along slices at the midplane, $z=0$, and
at the cylinder wall, $r=1$.  The general agreement between the two curves is
a manifestation of the near symmetry of $\ptwoD$. However, the curves behave
differently approaching the critical ring. Necessarily $\partial^2_z \ptwoD$
is even about $z=0$, since $\ptwoD$ is. There is no such constraint on
$\partial^2_r \ptwoD$ at $r=1$.  The important observation is that $\QtwoD <
\PtwoD < 0$, and hence that $|\PtwoD| < |\QtwoD|$. This means that the
pressure curvatures associated with just the meridional flow are unequal with
the ordering that promotes, rather than acts against, singularity formation.
While this ordering does not seem {\em a priori} obvious, it appears from
Fig.~\ref{fig:lines1}(b) to be a natural consequence of the conditions at the
wall and symmetry plane.  We will return to the importance of this ordering in
Sec.~\ref{sec:model}\ref{sec:connection}.

% From Matlab with file P2D:
% Q2D = DDPr_0 = -4.319227267398290e+07
% P2D = DDPz_0 = -2.855626370669210e+07
% ratio = a^2 = 1.512532350787225e+00
% sqrt(ratio) = a = 1.229850540019894e+00
%
% Qswirl = -8.679722747317683e+06
% Pswirl =  8.679722082134811e+06

%%%%%%%%%%%%%%%%%%  swirl text %%%%%%%%%%%%%%%%%% 

\subsection{Swirl pressure}

\begin{figure*}[h]
\centering
\includegraphics[width=1.0\linewidth]{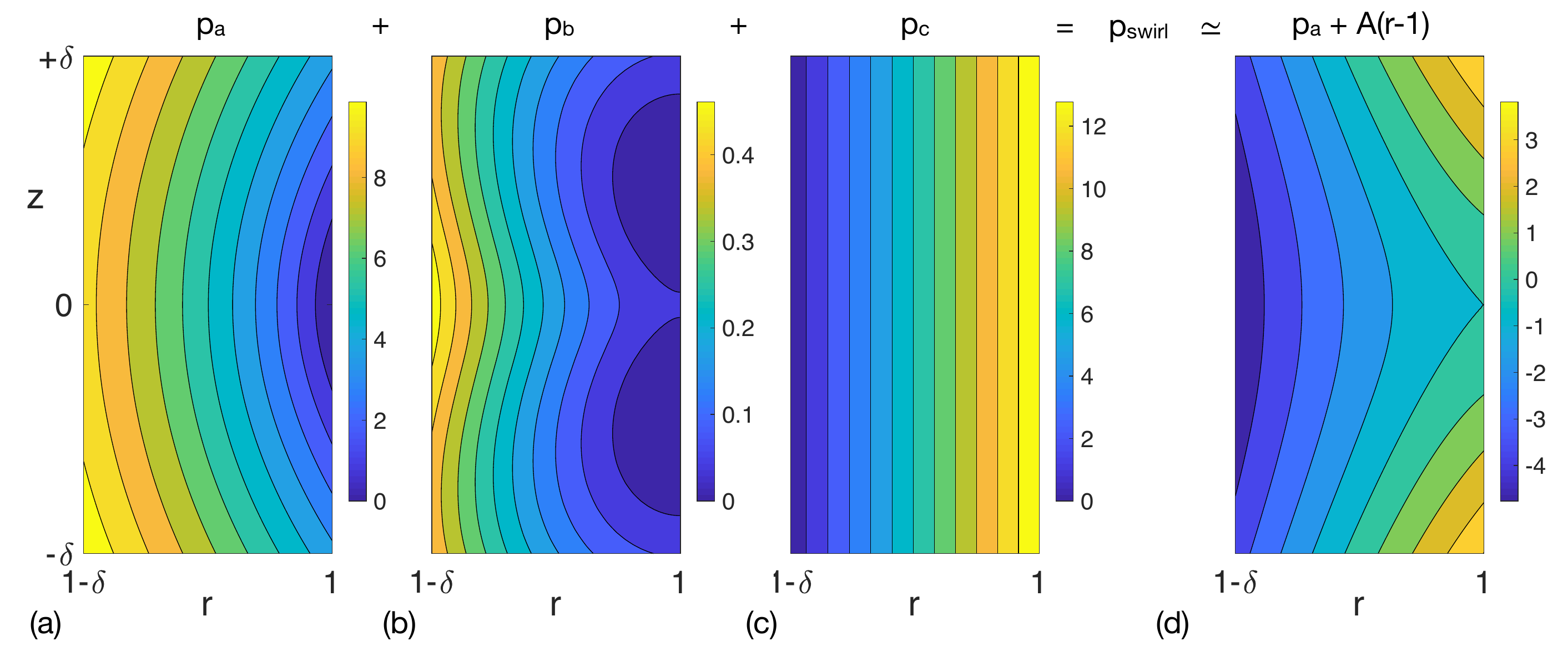}
\caption{Decomposition of the swirl pressure $\pswirl = p_a + p_b + p_c$
  visualised near the critical ring.  The component $p_a$ has a local minimum
  on the critical ring, but its second derivatives have opposite signs: $P_a >
  0 > Q_a$.  The range of $p_b$ is smaller than that of either $p_a$ or $p_c$
  and its variation along the cylinder wall $(r=1)$ is particularly weak in
  the region shown.  The component $p_c$ does not vary with $z$ by definition
  and it is nearly a linear function of $r$ in the region shown.  The
  right-most plot is $p_a + A (r-1)$, where $A = \atone{\partial_r p_c} =
  \atone{\langle u_\theta^2\rangle}$.  This field is barely distinguishable
  from $\pswirl$ shown in Fig.~\ref{fig:p_p2D_pswirl}.  It, and its curvatures
  $P_a$ and $Q_a$, are determined entirely by the swirl on the cylinder wall.
}
\label{fig:pa_pb_pc}
\end{figure*}

\begin{figure}[tbp]
\centering
\includegraphics[width=1.0\linewidth]{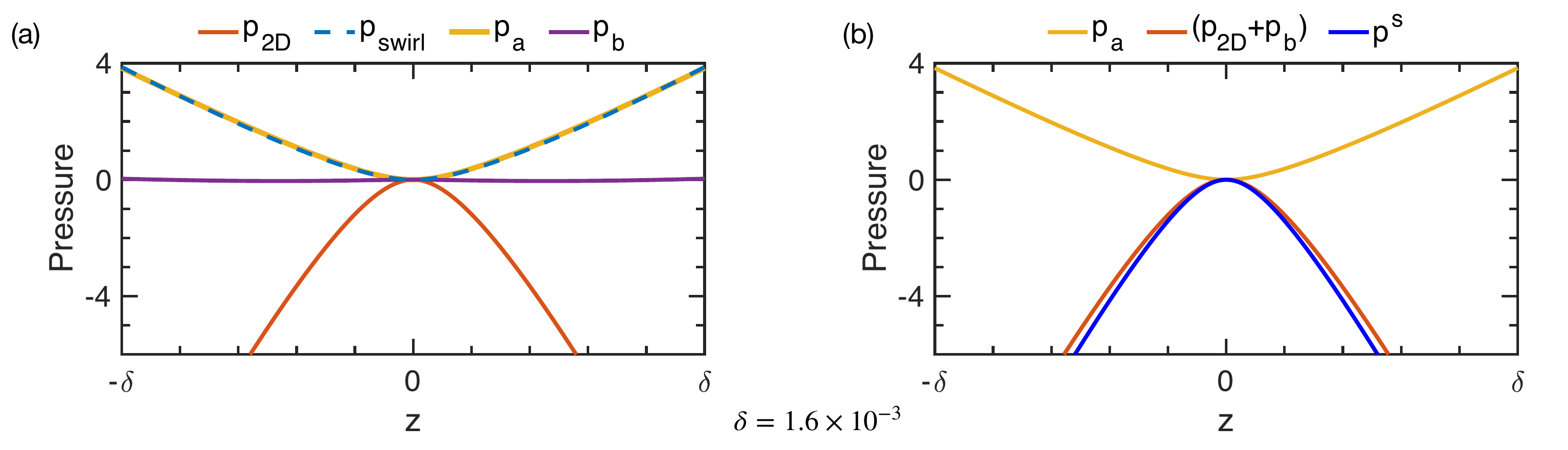}
\caption{
  Pressure components as function of $z$ along the cylinder wall $(r=1)$.
  Arbitrary constants are such that fields are all zero at $z=0$.  Since $p_c$
  does not vary with $z$, it is zero and not shown.
  (a) Along the cylinder wall near the critical ring $p_a$ and $\pswirl$ are
  nearly identical. Equivalently, $p_b = \pswirl - p_a$ is very small.  The
  axial curvature of $p_a$ is positive. The axial curvature of $p_b$ cannot be
  discerned in the plot, but it is negative on the critical ring.
  (b) Justification of the model closure. The two component fields with
  negative curvatures (adverse pressure gradients), $\ptwoD$ and $p_b$, are
  summed. (The graph of $\ptwoD+p_b$ is visually indistinguishable from that
  of $\ptwoD$ since $p_b$ is small.)  Also plotted is the field $\ptwoDmod$
  generated from the axial velocity of the Euler solution using
  \eqref{eq:ptwoDmod}.  This field from the symmetric approximation has a
  greater curvature magnitude and generates a larger adverse pressure gradient
  than the true field $\ptwoD+p_b$.
}
\label{fig:lines2}
\end{figure}

The swirl pressure $\pswirl$ not only maintains incompressibility of the flow,
it also confines the fluid within the cylinder wall.  Fully decoupling these
two effects is not achievable for flow in a cylinder, but we can mostly
separate them via the decomposition $p_{\rm swirl} = p_a + p_b + p_c$, where
\begin{subequations}
\begin{align}
  \nabla^2 p_a = 0, \quad 
  \atone{\partial_r p_a} & = \Bfluc \label{eq:Pa} \\
  \nabla^2 p_b = \Sfluc, \quad 
  \atone{\partial_r p_b} & = 0 \label{eq:Pb} \\
  \nabla^2 p_c = \Smean, \quad 
  \atone{\partial_r p_c} & = \Bmean \label{eq:Pc} 
\end{align}
\label{eq:pa_pb_pc}
\end{subequations}
where $\langle \cdot \rangle$ denotes axial mean and tilde denotes axial
fluctuations.  These pressure components are plotted in
Figs.~\ref{fig:pa_pb_pc} and \ref{fig:lines2}(a).  We also decompose the
pressure curvatures, $\Pswirl = P_a + P_b + P_c$ and $\Qswirl = Q_a + Q_b +
Q_c$, with the obvious meanings.  (See Appendix A for details of this
decomposition as well as relationships that hold for the component
curvatures.)

%%%%%%%%%%%%%%%%%%%%%%%%%%%%%%%%%%%%%%%%%%%%%%%%%%%%%%%%%%%%%%%%%%%%%%%%%%%%%

% From Matlab with file P2D:
%
% Qswirl = -8.679722747317683e+06
% Pswirl =  8.679722082134811e+06
% Pa     =  9.525754646424279e+06
% Pb     = -8.419045295319500e+05

%%%%%%%%%%%%%%%%%%%%%%%%%%%%%%%%%%%%%%%%%%%%%%%%%%%%%%%%%%%%%%%%%%%%%%%%%%%%%

The most significant fact from the decomposition is best seen in
Fig.~\ref{fig:lines2}(a). Near the critical ring, the axial variation of
$\pswirl$ is given almost exclusively by the component $p_a$.  The positive
axial curvature of $\pswirl$ comes about from the $p_a$ component: $\Pswirl
\simeq P_a > 0$.  (The radial curvatures satisfy $\Qswirl \simeq Q_a < 0$; see
Table \ref{tab:curvatures}. We return to this shortly.)  The stress field
associated with the pressure $p_a$ exists throughout the fluid solely to
provide force at the wall necessary to confine the flow within the cylinder --
the accelerations it generates within the fluid have no effect on the
divergence of the flow field.  The pressure field $p_a$ is the essence of the
teacup effect near the critical ring -- axial variation of the swirl at the
cylinder wall necessitates a pressure whose radial gradient at the wall
confines the fluid and whose axial gradient then forces axial flow toward the
critical ring.

The pressure field $p_b$ exists within the fluid to accelerate the flow and
suppress divergences that would otherwise arise from spatial variations of
$u_\theta$.  This component is very weak near the critical ring: the range of
values for $p_b$ is small in Fig.~\ref{fig:pa_pb_pc} and the curve
corresponding to $p_b$ in Fig.~\ref{fig:lines2}(a) is nearly flat.  Its
curvatures have signs $P_b < 0 < Q_b$ meaning that it acts against singularity
formation. However, these curvatures are an order of magnitude smaller than
those of $p_a$, so the effect is very weak. (See Table \ref{tab:curvatures}.)
Further from the critical ring, $p_b$ makes a more substantial contribution to
the momentum balance, but this is not important to singularity formation.

The pressure component $p_c$ is easy to interpret physically.  It is the
axially-independent pressure field that would be generated in the pure swirl
flow $\sqrt{\langle u_\theta^2 \rangle}(r) \hat e_\theta$, whose speed at each
$r$ is the axial r.m.s.\ of $u_\theta$.  The radially-inward force $-\nabla
p_c(r)$ is such as to curve each circular streamline of this r.m.s.\ swirl
flow, both maintaining incompressibility and confining the fluid at the
cylinder wall.  While this pressure component is significant in the radial
momentum balance, it contributes minimally, if at all, to the singularity.  By
definition $p_c$ does not vary with $z$, so it does not enter the axial
momentum balance and $P_c = 0$.  While $Q_c$ is not zero, it is the smallest
of all component pressure curvatures (see Table~\ref{tab:curvatures}).  (I
suspect that $Q_c$ does not diverge at the singularity and hence plays no role
in the blowup. See Appendix A.)

The essential aspect is this: the pressure component $p_a$ is the mechanism by
which the swirl on the wall couples to the pressure field. It is the sole
pressure component driving singularity formation.  It is visually evident in
Fig.~\ref{fig:lines2}(a) that $p_a$ is responsible for the positive curvature
of $\pswirl$ along the axial direction, and hence the favourable axial
pressure gradient. It is less evident comparing Fig.~\ref{fig:pa_pb_pc}(a) to
Fig.~\ref{fig:p_p2D_pswirl}(c) that $p_a$ dictates the negative radial
curvature of $\pswirl$.  This is because $\atc{\nabla \pswirl} = 0$, while
$\atc{\nabla p_a} \ne 0$.  To account for this, in Fig.~\ref{fig:pa_pb_pc}(d)
we show $p_a + A(r-1)$, where $A = \atc{\partial_r p_c} = -\atc{\partial_r
  p_a}$.  The field $A(r-1)$ is a linear approximation to $p_c$ at $r=1$, and
from the contours in Fig.~\ref{fig:pa_pb_pc}(c), $p_c$ is nearly linear in the
region shown.  Comparing Fig.~\ref{fig:pa_pb_pc}(d) to
Fig.~\ref{fig:p_p2D_pswirl}(c) it becomes clear that near the critical ring
$\pswirl \simeq p_a + A(r-1)$.  Since the term $A(r-1)$ is linear, the
curvatures of $p_a + A(r-1)$ are dictated solely by those of $p_a$.

(Briefly, the issue here is directly related to the impossibility of
separating the interior problem from the boundary condition for the axial mean
in a cylinder \eqref{eq:Pc}.  This is only problematic in that $p_a$ does not
``look like'' $\pswirl$ because $p_a$ lacks the component of the pressure
field responsible for confining the axial mean flow and hence its gradient is
non-zero on the critical ring.  Conceptually though, even if one could
separate the interior from the boundary for the axial mean, it would not
necessarily be desirable to add the resulting field to $p_a$, other than for
visual comparisons, because $p_a$ is the fundamental field that alone is
responsible for the opposite-signed curvatures driving singularity
formation. The axial mean is not important.  This decomposition of the swirl
pressure is a rich problem that I will not discuss further other than to note
that in the Boussinesq system (Appendix B and Sec.~\ref{sec:model})
confinement and incompressibility can be fully separated.)

%%%%%%%%%%%%%%%%%%%%%%%%%%%%%%%%%%%%%%%%%%%%%%%%%%%%%%%%%%%%%%%%%%%%%%%%%%%%%

\section{One-dimensional model and closure}
\label{sec:model}

I will now use facts learned from the pressure decomposition to gain insight
into mechanics of the blowup scenario.  The preceding analysis describes in
detail the situation at one time instant, but does not address the persistence
of this mechanism as the flow evolves.  To do this I will examine a model
based on a primitive-variables formulation of the Euler equations on the
cylinder wall, with closure coming from our knowledge of how pressure is
determined from velocity.

\subsection{Background}

There is a rich literature on one-dimensional modelling of singularities in
inviscid flow. See \cite{Choi_etal_2017} for a recent summary.  For the
cylinder flow, LH propose the model~\cite{LH_PNAS,LH_MMS}
\begin{equation}
  \partial_t \omega + u \partial_z \omega = \partial_z \theta, \quad 
  \partial_t \theta + u \partial_z \theta = 0, \label{eq:LHmodel}
\end{equation}
with the identifications $\omega(z) \sim \atone{\omega_\theta}$, $\theta(z)
\sim \atone{u_\theta^2}$, and $u(z) \sim \atone{u_z}$. (We abuse notation, by
conflicting with usage elsewhere in the paper and by not strictly
distinguishing between model quantities and their full-flow counterparts.)
\eqsref{eq:LHmodel} are closed by determining $u$ from $\omega$ via the
Hilbert transform \eqref{eq:Hdef}
\begin{equation}
\partial_z u = H(\omega). \label{eq:LH_Hilbert}
\end{equation}
The model and closure are natural from a vorticity-formulation viewpoint.  The
model captures very well features of the teacup flow~\cite{LH_MMS} and
exhibits a finite-time singularity~\cite{Choi_etal_2017}.

Details arise in interpretation of the LH model and the model presented
below. These are mostly relegated to Appendix B.  The essential point is that
away from the cylinder axis $r=0$, the axisymmetric Euler equations with swirl
have the same structure as the inviscid 2D Boussinesq equations posed on a
half-plane (what I shall refer to simply as the Boussinesq system; see
Appendix B).  In particular, because the two systems are equivalent on the
wall where the singularity occurs, it is convenient to invoke the structure of
the simpler Boussinesq system when considering model closures.  In the
Boussinesq system, closure models for evolution on the boundary come via the
Hilbert transform.  In discussing the model below, I will continue to use the
language of the axisymmetric Euler equations with swirl, but will invoke the
equivalence to the Boussinesq system as needed to close the model.

Of the three variables that appear in the LH model \eqref{eq:LHmodel}, two of
them, $\omega$ and $u$, are related via the Hilbert transform.  One can ask --
what about the Hilbert transform of the third variable $\theta$? From
\eqref{eq:PPE} and \eqref{eq:Pa} we have that
\begin{equation}
  H(\theta)
  = H(b)
  = H(\Bmean + \Bfluc)
  = H(\Bfluc)
  = H(\atone{\partial_r p_a})
  = -\atone{\partial_z p_a}. 
\label{eq:Hpa1}
\end{equation}
We have used linearity of $H$ and $H(\Bmean) = H({\rm const}) = 0$. The final
equality is exact, with the understanding that we are invoking the equivalence
to the Boussinesq system. (See Appendix B.)  Hence the Hilbert transform of
$\theta$ is, uniquely, the axial gradient of the pressure field $p_a$ on the
boundary. This is the unique physical meaning of $H(\theta)$.  Hence, even if
one did not set out to study the Euler equations in a primitive-variable
formation, one is lead to a decomposition of the pressure field just in
seeking to understand the meaning of $H(\theta)$.  It is important that the
variable $\theta$ in the LH model is equivalent to the axial gradient of
$p_a$.  For any model to capture the correct singularity mechanism, it must
capture $p_a$. The LH model does.  This helps to explain why the model can so
successfully capture the singularity using only variables on the cylinder
wall.

\subsection{Primitive-variable model}

The preceding suggests a different approach to closure -- working in a
primitive-variable formulation and obtaining pressure by Hilbert transform.
In the notation of this section, the Euler equations for the axial and swirl
flow on the wall are (exactly)
\begin{subequations}
\begin{align}
  \partial_t u + u \partial_z u & = -\partial_z p 
  = -\partial_z \ptwoD -\partial_z p_a -\partial_z p_b,
  \label{eq:Euler_wall_u}  \\
  \partial_t \theta + u \partial_z \theta & = 0. 
  \label{eq:Euler_wall_theta}  
\end{align}
\label{eq:Euler_wall}
\end{subequations}
Recall that $\partial_z p_c = 0$. 

These equations can be closed with a single modelling assumption that can be
justified from the simulation data.  First recall that near the critical ring
the contours of the meridional pressure $\ptwoD$ are nearly circular arcs
centred on the critical ring (Fig.~\ref{fig:p_p2D_pswirl}(b)).  If we assume
that $\ptwoD$ is exactly rotationally symmetric about the critical ring in the
meridional plane, (hence that its contours are exactly circular arcs), then
the meridional pressure is expressible just from source term $\StwoD$
evaluated on the wall. From this the associated axial pressure gradient is
\begin{equation}
-\partial_z \ptwoDmod =  \frac{2}{z} \int_0^z z' (\partial_z u)^2 \, dz',
\label{eq:ptwoDmod}
\end{equation}
See Appendix C for details. The superscript $s$ distinguishes this model
symmetric meridional pressure from the true meridional pressure gradient.  The
single modelling approximation we make is to replace the actual adverse
pressure gradients $-\partial_z \ptwoD - \partial_z p_b$ in
\eqref{eq:Euler_wall} by the symmetric pressure gradient $-\partial_z
\ptwoDmod$:
\begin{eqnarray}
  -\partial_z \ptwoD - \partial_z p_b 
& \rightarrow & 
  -\partial_z \ptwoDmod.
\label{eq:model_assump}
\end{eqnarray}
I address the validity of this approximation below. 

The favourable pressure gradient $-\partial_z p_a$ is given by the Hilbert
transform
\begin{equation}
-\partial_z p_a =  H(\theta).
\label{eq:Hpa2}
\end{equation}
This is an exact statement with the appropriate interpretation in terms of the
Boussinesq system and requires no other assumptions. 

Thus, we arrive at the model
\begin{subequations}
\begin{align}
& \partial_t u + u \partial_z u = -\partial_z \ptwoDmod -\partial_z
  p_a \label{eq:Pmodel_a} \\
  & \partial_t \theta + u \partial_z \theta =  0, \label{eq:Pmodel_b}
\end{align}
\label{eq:Pmodel}
\end{subequations}
where $-\partial_z \ptwoDmod$ is given by expression \eqref{eq:ptwoDmod} and
depends only on the axial flow $u$; $-\partial_z p_a$ is given by expression
\eqref{eq:Hpa2} and depends only on $\theta$, the square of the swirl.

\begin{figure}[h]
\centering
\includegraphics[width=1.0\linewidth]{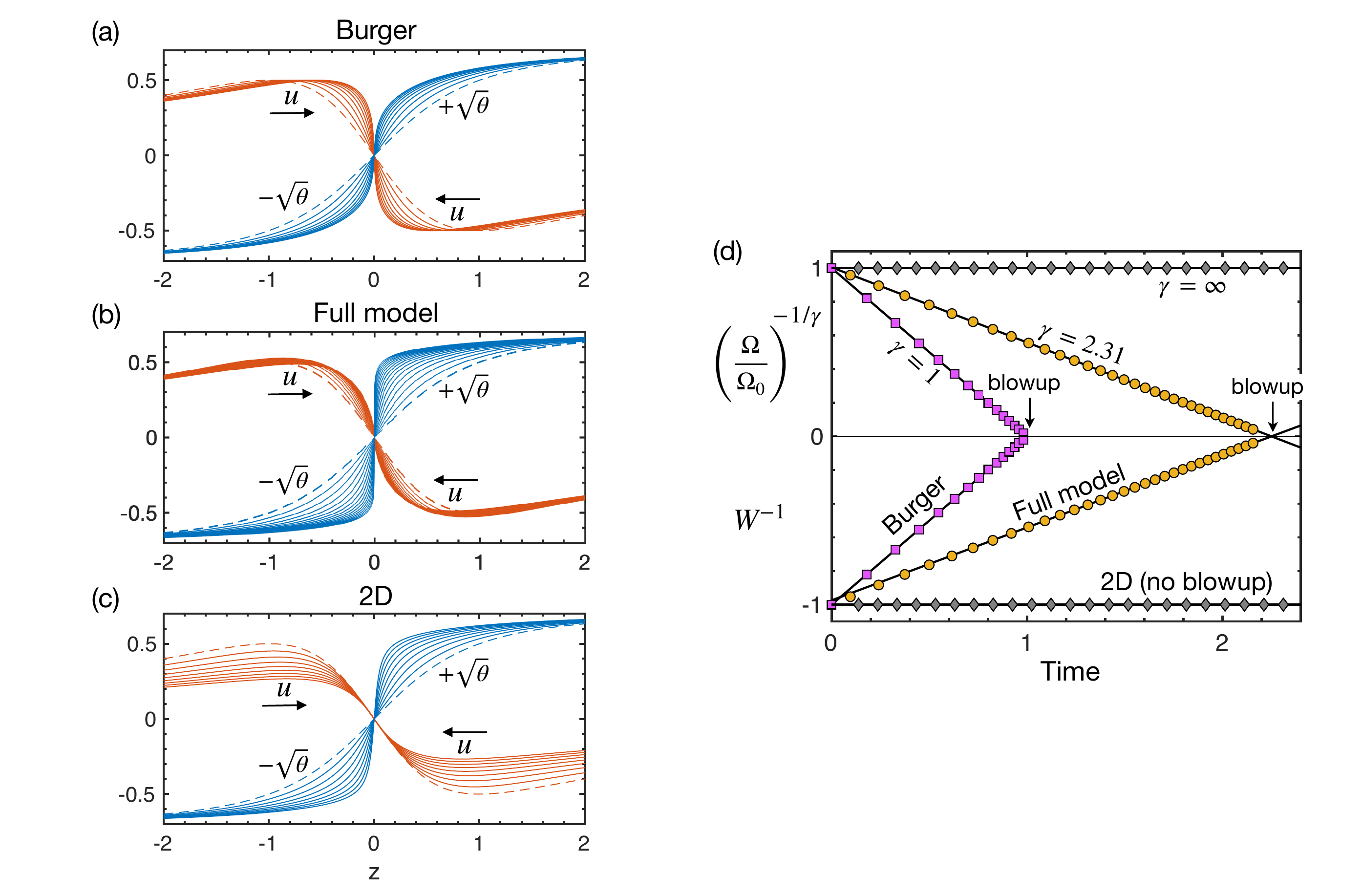}
\caption{Model simulations in three cases: (a) Burger's equation (no
  pressure), (b) the full model system \eqref{eq:Pmodel} with both pressure
  gradients $-\partial_z \ptwoDmod$ and $-\partial_z p_a$, and (c) the model
  with only the meridional (2D) pressure gradient $-\partial_z \ptwoDmod$.
  Plotted are representative snapshots during the evolution.  All start from
  the same initial condition (dashed curves).  The $\theta$ variable is
  plotted as $\pm \sqrt{\theta}$, since this corresponds to the swirl velocity
  $u_\theta$ on the cylinder wall. Arrow indicate the flow direction of the
  axial velocity $u$.  The slopes of these curves at $z=0$ are the velocity
  gradients $W$ and $\Omega$.
  (d) Time evolution of velocity gradients, where gradient are compensated
  based on the divergences in \eqref{eq:sol}.  Points are shown at
  representative times, not every time step.  In the Burger case, simulation
  data agrees well with the known exponent $\gamma=1$ and blowup time $T=1$.
  For the full model, $\gamma = 2.31$ has been estimated from the data (see
  text). The near linearity of the compensated data with the common
  extrapolated zero crossing support a blowup at time $T \simeq 2.250$.  With
  meridional pressure only (2D), $W$ is constant, $\gamma = \infty$, and there
  is no blowup.  }
\label{fig:u_theta_Omega_W}
\end{figure}

The results from simulations of this model are shown in
Fig.~\ref{fig:u_theta_Omega_W}. The initial condition is
\begin{equation}
u(z,t=0) = - \frac{z}{1 + z^2}, \quad 
\theta(z,t=0) = \frac{1}{2} \frac{z^2}{1 + z^2},
\end{equation}
where the factor $1/2$ is included in $\theta$ because with it the solution
almost immediately exhibits scaling behaviour.

Consider first just the case labelled ``full model'' -- meaning the model as
written in \eqref{eq:Pmodel}. The dynamics is illustrated in
Fig.~\ref{fig:u_theta_Omega_W}(b) with snapshots of the axial velocity $u$,
and the swirl velocity $\pm \sqrt{\theta}$.  The slopes of these curves at
$z=0$ are $W$ and $\Omega$, the velocity gradients on the critical ring. As
the system evolves in time, these gradients steepen from the incoming axial
flow.

To establish that the gradients blow up in finite time, we plot
$(\Omega/\Omega_0)^{-1/\gamma}$ and $W^{-1}$ as a function of time in
Fig.~\ref{fig:u_theta_Omega_W}(d).  Recall the form of the divergence given in
Eqs. \eqref{eq:sol} and note that the initial condition has $W_0=W(0)=-1$.
The linearity of these data, together with the common extrapolated zero
crossing at $T \simeq 2.250$, is strong evidence that $W$ and $\Omega$ blow up
in finite time.

The plot of $(\Omega/\Omega_0)^{-1/\gamma}$ requires a value for $\gamma$. This
can be estimated from the simulation data in two ways. First, the slope of
$W^{-1}$ versus $t$ in Fig.~\ref{fig:u_theta_Omega_W}(d) gives an estimate
of $1/\gamma$.  A least-squares fit of data over the range $1 \le t \le 2$
gives $\gamma \simeq 2.31$. The best-fit line is plotted.  A second estimate
of $\gamma$ is the value that minimises the residual error of a least squares
fit of $(\Omega/\Omega_0)^{-1/\gamma}$ versus $t$. Using the fitting range $1
\le t \le 2$, the minimum residual error is obtained for $\gamma \simeq 2.31$,
the same value to three digits of accuracy. The plot of
$(\Omega/\Omega_0)^{-1/\gamma}$ in Fig.~\ref{fig:u_theta_Omega_W}(d) uses
$\gamma = 2.31$ and the corresponding best-fit line is shown.  Note this
exponent is not very different from the value $\simeq 2.46$ obtained by LH for
the full Euler simulation.

Before discussing the implications of the model singularity for full Euler
flow, I want to return to the two special cases introduced in
Sec.~\ref{sec:illustrative}. The model contains both. Dropping all the
pressure terms, the model \eqref{eq:Pmodel} reduces to Burger's equation
together with the advection of the passive scalar $\theta$.  This system has a
finite-time singularity with divergences given in \eqref{eq:Burger_sol}. This
singularity is shown in Fig.~\ref{fig:u_theta_Omega_W}(a) and (d). The data
have been obtained from a computer simulation that differs from the full model
simulation only in the non-evaluation of the pressure terms within the
computer code.

The second special case corresponds to the situation with equal axial and
radial pressure curvatures on the critical ring: $P=Q$.  For the model,
$-\partial_z \ptwoDmod$ is obtained under the assumption that the pressure
field $\ptwoD$ is exactly rotationally symmetric in the meridional plane about
the critical ring (that the contours in Fig.~\ref{fig:p_p2D_pswirl}(b) are
exactly circular arcs). This assumption implies that the the axial and radial
curvatures are equal.  Hence dropping only $-\partial_z p_a$ from the model
\eqref{eq:Pmodel}, but keeping $-\partial_z \ptwoDmod$ gives this special
case.  The dynamics are shown Fig.~\ref{fig:u_theta_Omega_W}(c) and (d).  Here
there is strong deceleration of the axial flow, as seen by the ordering of the
$u$ snapshots in Fig.~\ref{fig:u_theta_Omega_W}(c) compared with the other two
cases.  The gradient of $u$ at $z=0$ is constant: $W(t)=W_0$, as in
\eqref{eq:2D_sol}.  This system does not blow up.  Again the data have been
obtained from a computer simulation that differs from the full model
simulation only in the non-evaluation of the $p_a$ term within the computer
code.

The model captures the interplay between inertia and pressure on the cylinder
wall.  When no stresses are included, the equations blow up in a Burger's
singularity and when only the stress associated with the two dimensional
meridional saddle is included, there is no blowup. In both these cases swirl
is advected as a passive scalar, leading to vorticity blowing up in the
Burger's case and exponentially growing vorticity in the non-blowup case.
Between these two is the case of interest, where the stress from the
confinement of swirl on the wall is included. Crucially, here swirl is not a
passive scalar -- as it is advected toward the critical ring by the axial
velocity, it generates increasingly large positive pressure curvature such
that the total pressure gradient is insufficient to decelerate in incoming
flow. Inertia overwhelms pressure gradients and blowup occurs.  The model
clearly shows how the teacup effect from the wall swirl is able to drive a
finite-time singularity.

\subsection{Connection to Euler}
\label{sec:connection}

It remains to relate the model to the full Euler flow.
The approximation made in the model closure \eqref{eq:model_assump}
encapsulates the difference between the two, and so I begin with this.

Recall the exact Euler equations on the cylinder wall \eqref{eq:Euler_wall},
where the axial pressure gradient $-\partial_z p$ separates into contributions
$-\partial_z \ptwoD$, $-\partial_z p_b$ and $-\partial_z p_a$.  These pressure
fields are plotted in Fig.~\ref{fig:lines2}(b) for the Euler solution at the
standard time considered in this paper.  The two components with negative
curvatures have been combined into $\ptwoD + p_b$. The corresponding adverse
pressure gradient $-\partial_z (\ptwoD + p_b)$ produces outward force
decelerating the axial flow approaching the critical ring.  Also plotted is
$\ptwoDmod$, the pressure obtained from \eqref{eq:ptwoDmod} using the actual
Euler flow on the cylinder wall.  One sees that $\ptwoDmod \le \ptwoD + p_b
\le 0$, with equality only on the critical ring.  This implies that in the
vicinity of the critical ring $|\partial_z \ptwoDmod | \ge |\partial_z \ptwoD
+ \partial_z p_b|$, meaning that the adverse pressure gradient $-\partial_z
\ptwoDmod$ based on the symmetry assumption provides more deceleration to the
incoming flow than the actual adverse pressure gradient $-\partial_z (\ptwoD +
p_b)$.

This justifies the closure approximation \eqref{eq:model_assump}, where the
actual pressure fields acting against singularity formation are replaced by a
field that {\em acts more strongly against singularity formation}.  In other
words, the closure approximation suggests that the model should be {\em less
  liable to blow up} than the full Euler equations.  This is important if we
want to draw inferences about singularities in the Euler equations from
singularity formation in the model. We want to know that we have not, at least
not in an obvious way, introduced a singularity mechanism through the model
closure.

Another way to view the connection between the model and the Euler equations
is via the velocity gradient dynamics on the critical ring.
Taking the $z$-derivative of model equations \eqref{eq:Pmodel} and evaluating
at $z=0$ gives
\begin{equation}
\dot W = - P_a,  
\quad \dot \Omega + W \Omega = 0  \quad  \mbox{(model)}.
\label{eq:Wdot_model}
\end{equation}
By construction, the curvature of the symmetric pressure $\ptwoDmod$ exactly
balances the inertial nonlinearity on the critical ring, leaving only the
pressure curvature $P_a = -\atzero{\partial_z^2 p_a}$ driving the velocity
gradient $W$.

For actual Euler flow we have instead the inequality
\begin{equation}
\dot W < - P_a, 
\quad \dot \Omega + W \Omega = 0  \quad \mbox{(Euler)}.
\label{eq:Wdot_Euler}
\end{equation}
This follows from \eqref{eq:main} under the condition that $\QtwoD < \PtwoD +
2 P_b$.  This brings us back to the key observation seen in
Fig.~\ref{fig:lines1}(b), namely that the meridional pressure curvatures are
not exactly equal, $\PtwoD \ne \QtwoD$.  From the data in Table
\ref{tab:curvatures}, $\PtwoD$ and $\QtwoD$ are sufficiently different that
the inequality $\QtwoD < \PtwoD + 2 P_b$ holds ($2P_b$ is an order of
magnitude smaller than the difference between $\PtwoD$ and $\QtwoD$).  In fact
the inequality follows from the previous observation that $\ptwoDmod \le
\ptwoD + p_b \le 0$ with equality only on the critical ring.

The difference between the equality in \eqref{eq:Wdot_model} and the
inequality in \eqref{eq:Wdot_Euler} quantifies the previous point that the
closure approximation appears to be safe, in that it does not (obviously)
enhance singularity formation over that of Euler flow.  (The singularities
occur with $W \to -\infty$.)  For simplicity of discussion, throughout this
section I have not strictly distinguished between model and full-Euler
quantities.  Here it is essential to be clear.  Equations
\eqref{eq:Wdot_model} and \eqref{eq:Wdot_Euler} use the same symbols, but
apply to different (but closely related) systems -- \eqref{eq:Wdot_model}
holds for solutions of the model equations \eqref{eq:Pmodel}, while
\eqref{eq:Wdot_Euler} holds for solutions of the full Euler equations. More
specifically, \eqref{eq:Wdot_model} holds exactly by construction;
\eqref{eq:Wdot_Euler} holds by numerical observation of the Euler solution and
is presumed to hold up to the singularity time.

Although \eqref{eq:Wdot_model} and \eqref{eq:Wdot_Euler} do not establish a
rigorous relationship between the model and Euler flow, they nevertheless
reduce the mechanism for singularity formation, in both cases, to its most
basic form.  Within the Boussinesq analogy, $P_a$ is given by the same
function of wall swirl in both cases.  Using \eqref{eq:Hpa2} we have $P_a
\equiv \atzero{\partial_z^2 p_a} = -H(\partial_z \theta)(0)$. Depending on the
case, either $\theta$ comes from the solution of the model \eqref{eq:Pmodel}
or else from the swirl on the wall from Euler flow.  Referring to
\eqref{eq:main2}, for either system to blowup, the pressure curvature due to
swirl on the cylinder wall must diverge as $W^2$.  Specifically, we can relate
$W^2/\gamma$ to $P_a$ in \eqref{eq:Wdot_model} and \eqref{eq:Wdot_Euler} to
give
\begin{equation}
-\frac{H(\partial_z \theta)(0)}{W^2} \le \frac{1}{\gamma}.
\label{eq:select}
\end{equation}
Either system will blow up if the left-hand side remains bounded above
zero by any finite amount. 

In principle, \eqref{eq:select} provides a selection mechanism for the
exponent $\gamma$. It selects $\gamma$ sharply in the case of the model and
bounds $\gamma$ in the case of Euler flow. It is a global condition relating
the swirl everywhere on the wall to the velocity gradient on the critical
ring.  For the model, one can verify numerically that the axial velocity and
swirl evolve together such that $-H(\partial_z \theta)(0)/W^2$ gives the value
of $\gamma$. However, this is a triviality given the evidence of a singularity
already presented in Fig.~\ref{fig:u_theta_Omega_W}.  Other than numerical
simulations, I have been unable to find any convincing arguments or insights
into how a particular value of $\gamma$ is selected. I leave this for future
work.

%%%%%%%%%%%%%%%%%%%%%%%%%%%%%%%%%%%%%%%%%%%%%%%%%%%%%%%%%%%%%%%%%%%%%%%%%%%%%

\section{Conclusion}

The potential Euler singularity discovered by Luo and
Hou~\cite{LH_PNAS,LH_MMS} has significantly advanced our mathematical
understanding of finite-time singularities and it provides a concrete, easily
reproducible case to explore computationally.  Here I have sought to
understand this singularity from a mechanics point of view and from this gain
physical insights into why this particular flow configuration permits velocity
gradients to blow up in finite time.

The analysis focuses on the interplay between inertia and pressure.  A direct
connection is established between the singularity mechanism and flow
confinement. The pressure field at the heart of the teacup effect is present
solely to confine the rotating fluid within the cylinder; it is determined
only by the swirl on the cylinder wall and it plays no role in maintaining
incompressibility of the flow.  This field is responsible for unequal axial
and radial pressure curvatures on the critical ring.  This inequality of
pressure curvatures is precisely the condition needed for fluid inertia to
overwhelm the adverse pressure gradient on the cylinder wall and for velocity
gradients to blow up.

To understand how this scenario plays out, a new model has been proposed based
on a primitive-variable formulation of the Euler equations. The model
describes axial and swirl velocities on the cylinder wall, with closure coming
from the dependence of pressure on these velocities. For the swirl the
pressure is known exactly. For the axial velocity an approximation is made
that has a physical meaning and is supported by Euler simulations. This
approximation appears to be distinctly different from those used in related
models~\cite{LH_PNAS,LH_MMS,CKY_2015}.

The model captures the interplay between inertia and pressure gradients on the
cylinder wall and moreover is embedded in a broader class of problems.  In one
limit, there are no stresses acting and hence no deceleration of axial
flow. This leads to an easily understood Burger's singularity, accompanied by
vorticity blowup from the transport of swirl as a passive scalar.  At the
other limit, there is only the stress associated with the acceleration of flow
around a two dimensional saddle point. This leads to substantial deceleration
of the axial flow and no blowup.  Between these limits is the case that
includes both the stress due to the saddle point and the stress generated from
confinement of the swirl on the wall (the teacup effect). Swirl is then not a
passive scalar -- as it is advected toward the critical ring by the axial
velocity, it generates an increasingly large positive pressure curvature (and
associated favourable pressure gradient) such that the total pressure gradient
is insufficient to decelerate incoming flow. Velocity gradients blow up in a
singularity.

There is an important connection between this mechanism and other recent
popular models for singularity formation~\cite{LH_PNAS,LH_MMS,CKY_2015}. These
models contain two variables, vorticity and square swirl on the cylinder
wall. The Hilbert (or similar) transform of the vorticity is used to obtain
velocity. The Hilbert transform of the square swirl is, uniquely, the axial
gradient of the confining pressure at the core of the mechanism described
here. 

There are many future directions suggested by this work.  Pressure could
possibly provide physical insight into the role of the boundary in the rapid
growth of vorticity gradients shown by Kiselev and
\v{S}ver\'{a}k~\cite{Kiselev_Sverak_2014}.  The model closure proposed here
could be connected to the hyperbolic system studied by Kiselev and Tan
\cite{Kiselev_Tan_2018}. Along these same lines, to impart greater equivalence
between axial flow near the cylinder wall and radial flow near $z=0$, one
could simulate a cylindrical configuration with a no-penetration condition at
$z=0$.  The Euler simulations presented here are only for the specific case
(initial condition and cylinder aspect ratio) used by Luo and Hou and this
leaves open the question of how singularity formation in wall-bounded swirling
flows depends on these. One presumes that the scaling exponent $\gamma$
is independent of such factors, as long as blowup occurs, but this too is not
presently known since the selection mechanism for the exponent $\gamma$
remains open.  It would be highly desirable to investigate these issues and to
consider other geometries such as swirling flow within a sphere and to
understand the role of pressure in other configurations, such as anti-parallel
vortices \cite{Bustamante_Kerr}.  It seems likely that the blowup observed
numerically in the model is self-similar, but this is unknown at present, and
currently there is no proof of blowup in the model equations.  It should be
possible to develop precise theorems along the lines of Chae and collaborators
\cite{chae2008incompressible,chae2008blow,constantin2008singular,
  chae2010lagrangian, Chae_etal_2012} to address the specific pressure fields
described here. This could possibly lead to a new line of attack on proof of a
singularity in the Euler equations.  Finally, and most fundamentally, flow
confinement is key to the mechanism described here and hence the question
remains open as to whether an Euler solution can exhibit blowup in a
configuration without a pressure field originating from flow confinement.

% \dataccess{ The numerical Euler solution at time t=0.0031 and processed data
%   used in plotting the pressure fields are provided as electronic
%   supplementary material.  }

\begin{acknowledgments}
  
{This work was partially supported by a grant from the Simons Foundation
  (Grant number 662985, NG).}

{I am grateful to Guo Luo for pointing out an error in an earlier
  manuscript and for providing data with which the present simulations could
  be validated. I became interested in this problem during the IPAM program on
  the Mathematics of Turbulence and I thank IPAM for their support. }

\end{acknowledgments}

%%%%%%%%%%%%%%%%%%%%%%%%%%%%%%%%%%%%%%%%%%%%%%%%%%%%%%%%%%%%%%%%%%%%%%%%%%%%%

\section*{Appendix A: Pressure decomposition}

Here we provide details of the pressure decomposition and summarise the
relationships that exist between pressure curvatures on the critical ring.  In
component form, the Euler equations for axisymmetric flow with swirl are
\begin{subequations}
\begin{align}
\partial_t u_r + \hat u \cdot \hat \nabla u_r - \frac{u_\theta^2}{r} & = 
-\partial_r p \label{eq:Euler_r} \\
\partial_t u_\theta + \hat u \cdot \hat \nabla u_\theta
+ \frac{u_r u_\theta}{r} & = 0 \label{eq:Euler_t} \\
\partial_t u_z + \hat u \cdot \hat \nabla u_z & = -\partial_z p
\label{eq:Euler_z} 
\end{align}
\label{eq:Euler_coord} 
\end{subequations}
where $\hat u = (u_r, u_z)$ and $\hat \nabla = (\partial_r, \partial_z)$. 

Taking the divergence of the nonlinear terms gives the source term $\Source$
on the right-hand-side of the pressure Poisson equation
\begin{align*}
\Source = -\frac{1}{r} \partial_r \left( r \hat u \cdot \hat \nabla u_r \right)
  + \frac{1}{r}\partial_r u_\theta^2 
  - \partial_z \left( \hat u \cdot \hat \nabla u_z \right)
\end{align*}
The first and third terms are independent of the swirl velocity $u_\theta$,
while the middle term depends only on $u_\theta$. This leads us to define
\begin{align}
  \StwoD & =
  -\frac{1}{r} \partial_r \left( r \hat u \cdot \hat \nabla u_r \right)
  - \partial_z \left( \hat u \cdot \hat \nabla u_z \right), \quad
  \Sswirl =
  \frac{1}{r}\partial_r u_\theta^2. \label{eq:sources}
\end{align}
Thus the pressure Poisson equation, with boundary condition, is
\begin{align*}
\nabla^2 p = \Source = \StwoD + \Sswirl, \quad 
\left. \partial_r p \right\vert_{r=1} = \atone{u_\theta^2} = \Bound
\end{align*}
This allows for the pressure to be decomposed as $p = \ptwoD + \pswirl$, as
given in \eqref{eq:p2D_pswirl}. 

Then $\Sswirl$ and $\Bound$ can be further decomposed into axial mean and
fluctuating terms
\begin{align*}
\Sswirl = \Smean + \Sfluc, \quad
\Bound  = \Bmean + \Bfluc,
\end{align*}
where $\langle \rangle$ denotes axial mean, 
$$
\langle{f}\rangle(r) = \frac{1}{L} \int_0^L f(r,z) \, dz
$$
This allows for the swirl pressure to be decomposed as $p_{\rm swirl} = p_a +
p_b + p_c$, as given in \eqref{eq:pa_pb_pc}. 

For the velocity gradient dynamics we require the pressure Hessian $\nabla
(\nabla p)$. The pressure field satisfies $\partial_\theta p = 0$ everywhere.
Since $p$ is even in $z$, it also satisfies $\atzero{\partial_z p} = 0$.  Hence
at $z=0$ the pressure Hessian is
$$
\atzero{\nabla(\nabla p)}  =
\begin{bmatrix}
\atzero{\partial_r^2 p} & 0 & 0 \\
0 & \frac{1}{r}\atzero{\partial_r p}  & 0 \\
0 & 0 & \atzero{\partial_z^2 p}
\end{bmatrix}
$$
with the ordering of components $r, \theta, z$. 
On the critical ring, $\atc{\partial_r p} = 0$ since $\atc{b} = 0$, and the
pressure Hessian is 
$$
\atzero{\nabla(\nabla p)}  =
\begin{bmatrix}
Q & 0 & 0 \\
0 & 0 & 0 \\
0 & 0 & P 
\end{bmatrix}
$$
The Laplacian of $p$ is the trace of the Hessian, so on the critical ring
$\nabla^2 p = Q + P$.

From the decomposition, the curvatures for the component fields obey
\begin{align}
P & = \PtwoD + \Pswirl = \PtwoD +  P_a + P_b + P_c, \\
Q & = \QtwoD + \Qswirl = \QtwoD + Q_a + Q_b + Q_c.
\end{align}
There are relationships that hold for the component pressure curvatures on the
critical ring. From 
$
\atc{\nabla^2 \pswirl} = \atc{\Sswirl} = 
\atc{\frac{1}{r}\partial_r u_\theta^2} = 0$, 
we have immediately $\Pswirl + \Qswirl = 0$ leading to \eqref{eq:PQ_decomp}. 

Less trivial relationships hold for the decomposition of $\pswirl$ into $p_a +
p_b + p_c$. The reason is that $-\atc{\partial_r p_a} = \atc{\partial_r p_c}
= \Bmean \ne 0$. Hence these terms appear in the pressure Hessian for $p_a$
and $p_c$.  From \eqref{eq:Pa}, $\atc{\nabla^2 p_a} = 0$, giving
\begin{equation}
P_a + \atc{\partial_r p_a} + Q_a = 0.
\end{equation}
From \eqref{eq:Pb} and \eqref{eq:Pc}, $\atc{\nabla^2 (p_b +
  p_c)} = 0$, giving
\begin{equation}
P_b + Q_b + Q_c + \atc{\partial_r p_c} = 0,
\end{equation}
where we have used that $P_c = 0$. Note that while the second derivatives of
pressure blowup at the singularity, the first derivatives do not. This is
because $-\atc{\partial_r p_a} = \atc{\partial_r p_c} = \Bmean = \langle
u_\theta^2 \atone{\rangle}$, and $u_\theta$ does not blowup. Hence, close to
the singularity 
\begin{equation}
P_a + Q_a  \simeq 0 \qquad P_b + Q_b + Q_c  \simeq 0,
\label{eq:PaQaPbQbQc}
\end{equation}
where approximately zero means here that the sums are not diverging even
though the individual terms are. The simulations suggest that $Q_c$ does not
blowup at the singularity and can be dropped from \eqref{eq:PaQaPbQbQc}.  This
is reasonable since $Q_c + \atc{\partial_r p_c} = \partial_r \langle
u_\theta^2 \atone{\rangle}$, and so for $Q_c$ to blowup, the gradient of the
axial mean must blowup. There is possibly an easy demonstration that this
cannot occur.

%%%%%%%%%%%%%%%%%%%%%%%%%%%%%%%%%%%%%%%%%%%%%%%%%%%%%%%%%%%%%%%%%%%%%%%%%%%%%

\section*{Appendix B. Connection to 2D Boussinesq system and Hilbert transform}

There is a well-known relationship between axisymmetric flow with swirl and
two-dimensional thermal convection in the inviscid Boussinesq approximation.
See in particular \cite{Majda_Bertozzi,Choi_etal_2017}.  The Euler equations
for axisymmetric flow with swirl \eqref{eq:Euler_coord} can be recast as,
\begin{subequations}
\begin{align}
\partial_t \hat u + \hat u \cdot \hat \nabla \hat u & = 
- \hat \nabla p + \frac{(r u_\theta)^2}{r^3} \hat e_r, \label{eq:cyl1} \\
\partial_t (r u_\theta) + \hat u \cdot \hat \nabla (r u_\theta) 
& = 0,  \label{eq:cyl2} 
\end{align}
\label{eq:cyl}
\end{subequations}
where \eqref{eq:cyl1} is a vector equation for the meridional flow $\hat u =
(u_r, u_z)$ obtained by combining \eqref{eq:Euler_r} and
\eqref{eq:Euler_z}. The centripetal acceleration term has been written in
terms of $(r u_\theta)^2$ and moved to the right-hand side.  Equation
\eqref{eq:cyl2} is just a reformulation of \eqref{eq:Euler_t} into a form that
expresses conservation of $r u_\theta$ as it is advected as a passive scalar
by the meridional flow.  Letting $\theta = (r u_\theta)^2$, the equations take
the simple from
\begin{subequations}
\begin{align}
\partial_t \hat u + \hat u \cdot \hat \nabla \hat u & = 
- \hat \nabla p + \frac{\theta}{r^3} \hat e_r, \label{eq:cyl1_mytheta} \\
\partial_t \theta + \hat u \cdot \hat \nabla \theta & = 0.  
\end{align}
\end{subequations}
In this form, $\theta$ can be viewed as providing a radial driving to the
meridional flow. (It should be emphasised, however, that the $\theta$-term
in \ref{eq:cyl1_mytheta} comes from inertia seen in cylindrical
coordinates. This term is not associated with stresses acting within the
fluid.)

For the inviscid 2D Boussinesq system, consider two-dimensional flow $u(x,y) =
u_x(x,y) \hat e_x + u_y(x,y) \hat e_y$ in the region $y \ge 0$.  In the
Boussinesq approximation, one allows for density variations within the fluid
due to thermal expansion from temperature variations.  Gravity acts on the
density field, here pointing in the $-\hat e_y$ direction, and the governing
equations are
\begin{subequations}
\begin{align}
  \partial_t u + u \cdot \nabla u & = - \nabla p - \rho \hat e_y
\label{eq:Boussinesq1}  \\
\partial_t \rho + u \cdot \nabla \rho & = 0 
\label{eq:Boussinesq2}  
\end{align}
\end{subequations}
where $\rho$ represents the density variation relative to some background
density.  Eq.~\eqref{eq:Boussinesq1} describes momentum balance, while
Eq.~\eqref{eq:Boussinesq2} describes the advection of the density field as a
passive scalar.  Just as viscosity is zero, thermal diffusivity is zero in
this system (both molecular effects are omitted).

\begin{figure}[h]
\centering
\includegraphics[width=1.0\linewidth]{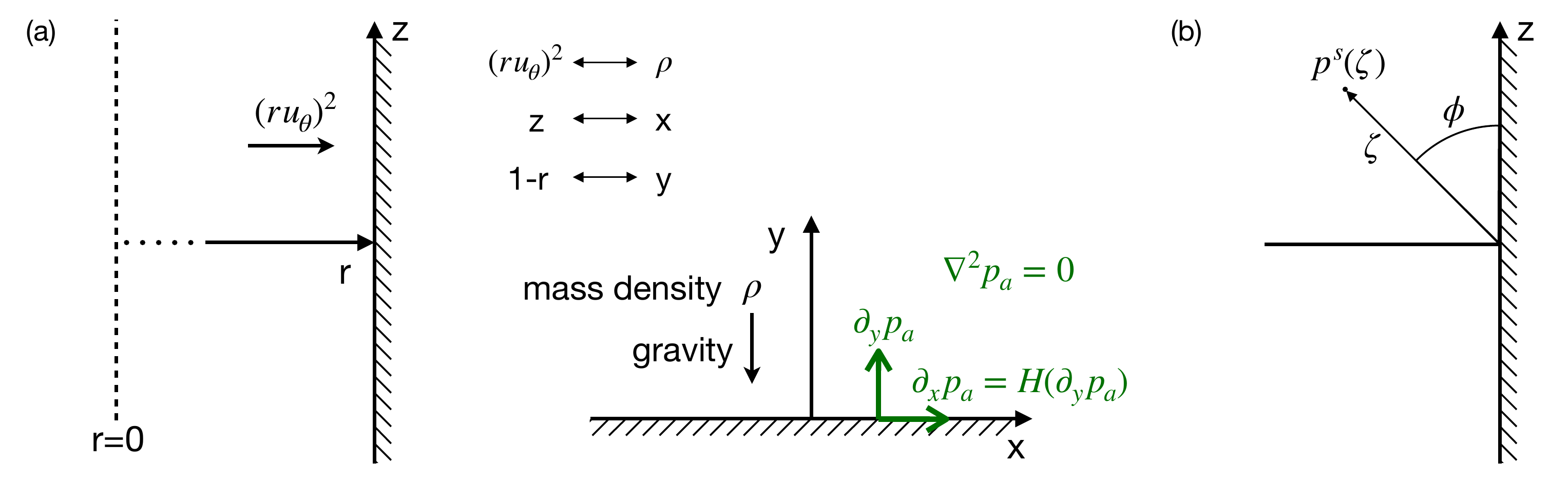}
\caption{(a) Correspondence between axisymmetric flow with swirl away from the
  axis and the inviscid 2D Boussinesq system. Expressions in green illustrate
  that if $p_a$ is a harmonic function in the upper half plane then its
  tangential derivative is the Hilbert transform of its normal derivative.  (b)
  Coordinates for symmetric pressure field $\ptwoDmod$. }
\label{fig:appendix}
\end{figure}

The correspondence between the two systems is illustrated in
Fig.~\ref{fig:appendix}(a). Quoting from \cite[p. 187]{Majda_Bertozzi}, ``we
see that the 2D Boussinesq equations are formally identical to the equations
for 3D axisymmetric, swirling flows provided that we evaluate all external
variable coefficients'' $\dots$ ``at $r=1$. Thus away from the axis of
symmetry $r=0$ for swirling flows, we expect the qualitative behaviour of the
solutions for the two systems of equations to be identical.''

The advantage of the cylindrical system is it is straightforward to simulate
numerically.  The advantage of the Boussinesq system is that it is
mathematically simpler.  We may consider either system on a periodic or on an
infinite domain in the axial, $z$, or horizontal, $x$, direction. The infinite
case is the simplest to consider conceptually and it what I will mean by the
Boussinesq system.

This brings us to the Hilbert transformation.  Consider a harmonic function
$\phi$ in the upper half plane. In our case $\phi$ will be the pressure
component $p_a$ associated with the boundary swirl, or boundary density for
the Boussinesq system.  The Hilbert transform of the normal derivative of
$\phi$ along $y=0$ is the tangential derivative of $\phi$ along $y=0$.  This
gives the fundamental relationship between the swirl and axial pressure
gradient of $p_a$ on the cylinder wall. Hence, the Hilbert transform appears
in Sec.~\ref{sec:model}. A minus sign arises in \eqref{eq:Hpa1} and
\eqref{eq:Hpa2} because $y$ analogous to $1-r$, so $\partial_y = -\partial_r$.
Concretely, the Hilbert transform of a function $f(x)$ is defined as
\cite[p. 173]{Majda_Bertozzi}
\begin{equation}
H(f)(x) = \frac{1}{\pi} PV \int_{-\infty}^\infty \frac{f(x')}{x - x'} dx',
\label{eq:Hdef}
\end{equation}
where $PV$ denotes principle value.

%%%%%%%%%%%%%%%%%%%%%%%%%%%%%%%%%%%%%%%%%%%%%%%%%%%%%%%%%%%%%%%%%%%%%%%%%%%%%

\section*{Appendix C. Meridional pressure with symmetry assumption}

Assume that in a meridional plane a pressure field $\ptwoDmod$ is exactly
rotationally symmetric about the critical ring (that the contours in
Fig.~\ref{fig:p_p2D_pswirl}(b) are exactly circular arcs).  This assumption
necessarily requires invoking the Boussinesq analogy because such a symmetry
is impossible within a cylinder. As elsewhere, I nevertheless use here the
language of the axisymmetric Euler equations with swirl.  Let $(\zeta,\phi)$
be polar coordinates centred on the critical ring as shown in
Fig.~\ref{fig:appendix}. Then $\ptwoDmod$ is a function only of $\zeta$.  We
assume $\ptwoDmod$ is determined by a pressure Poisson equation $\nabla^2
\ptwoDmod = \StwoDmod$, where the source $\StwoDmod$ must also be rotationally
symmetric and hence only a function of $\zeta$. Considering the ray $\phi=0$
and identifying $\zeta$ with the positive $z$ axis, we set $\StwoDmod =
\StwoD(r=1,z)$, where $\StwoD(r=1,z)$ is the source term for Euler flow
\eqref{eq:sources} evaluated on the cylinder wall.  Straightforward
calculation gives $\StwoD(r=1,z) = -2(\partial_z u)^2$, where $u =
u_z(r=1,z)$, from which
\begin{equation}
  \nabla^2 \ptwoDmod =
\frac{1}{z} \partial_z \left(z
\partial_z \ptwoDmod \right) = -2(\partial_z u)^2.
\end{equation}
Integrating this once gives \eqref{eq:ptwoDmod}.

%%%%%%%%%%%%%%%%%%%%%%%%%%%%%%%%%%%%%%%%%%%%%%%%%%%%%%%%%%%%%%%%%%%%%%%%%%%%%

\section*{Appendix D. Numerical simulations}

The Euler equations have been simulated in the vorticity-streamfunction
formulation as given by Eqs.~(2) in~\cite{LH_PNAS}. The essential difference
between the simulations here and those of LH~\cite{LH_PNAS,LH_MMS} is that
here a fixed computation grid is used. A Fourier pseudospectral representation
is used in $z$ with dealiasing given by Hou and Li~\cite{Hou_Li_2007}. A
Chebychev grid is used in $r$ with no dealiasing. Fourth-order Runge-Kutta
time stepping is used with an adaptive time step such that the CFL number is
less than 0.2.  Exploiting the separation in the Fourier representation, the
Poisson problem for the streamfunction is solved directly.  Solving similar
Poisson problems, pressure fields are computed in a post-processing step.

For all results reported the computation grid has 769 radial points for $r \in
[0,1]$ and 2048 axial points for $z \in [0,L/4)$.  At time $t=0.0031$
  simulations produce a vorticity maximum $\|\omega\|_\infty = 1.54276898
  \times 10^5$, agreeing to about 8 digits of precision with the value
  $\|\omega\|_\infty = 1.54276901 \times 10^5$ from simulations by Luo and Hou
  (private communication).

The simulations of the model equations \eqref{eq:Pmodel} are mostly
straightforward. The $z$ coordinate is mapped to $x \in (-1,1)$ via $\lambda
\pi z/2 = \tan(\pi x/2)$, where the parameter $\lambda = 8$ is used to
increase the resolution near $z=0$. The integrals \eqref{eq:ptwoDmod} and
\eqref{eq:Hpa2} are computed by quadrature (taking into account the symmetry of
the solution). Derivatives are computed spectrally with 2048 equally spaced
grid points in $x$.  Fourth-order Runge-Kutta time stepping is used with a
time step such that the CFL number is fixed at 0.2.

%%%%%%%%%% Insert bibliography here %%%%%%%%%%%%%%

\vskip2pc

% \bibliographystyle{RS}
% \bibliography{paper}

\begin{thebibliography}{99}

\bibitem{Einstein}
Einstein A. 1926  Die Ursache der M{\"a}anderbildung der Flu{\ss}l{\"a}ufe und
  des sogenannten Baerschen Gesetzes. {\em Naturwissenschaften} \textbf{14},
  223--224.

\bibitem{LH_PNAS}
Luo G, Hou TY. 2014a  Potentially singular solutions of the 3D axisymmetric
  {E}uler equations. {\em Proc Natl Acad Sci USA} \textbf{111}, 12968--12973.

\bibitem{LH_MMS}
Luo G, Hou TY. 2014b  Toward the {Finite-Time} Blowup of the {3D} Axisymmetric
  {E}uler Equations: A Numerical Investigation. {\em Multiscale Model Sim.}
  \textbf{12}, 1722--1776.

\bibitem{LH_Review}
Luo G, Hou TY. 2019  Formation of Finite-Time Singularities in the 3D
  Axisymmetric Euler Equations: A Numerics Guided Study. {\em SIAM Review}
  \textbf{61}, 793--835.

\bibitem{chae_Tsai_2015}
Chae D, Tsai TP. 2015  Remark on {Luo-Hou's} ansatz for a self-similar solution
  to the 3D {E}uler equations. {\em J. Nonlinear Sci.} \textbf{25}, 193--202.

\bibitem{Sperone_2017}
Sperone G. 2017  Further Remarks on the {Luo-Hou's} Ansatz for a Self-similar
  Solution to the {3D} {E}uler Equations. {\em J. Nonlinear Sci.} \textbf{27},
  1325--1338.

\bibitem{chae2008incompressible}
Chae D. 2008a  Incompressible Euler Equations: the blow-up problem and related
  results. {\em Handbook of Differential Equations: Evolutionary Equations}
  \textbf{4}, 1--55.

\bibitem{chae2008blow}
Chae D. 2008b  On the blow-up problem for the axisymmetric 3D Euler equations.
  {\em Nonlinearity} \textbf{21}, 2053.

\bibitem{constantin2008singular}
Constantin P. 2008  Singular, weak and absent: Solutions of the Euler
  equations. {\em Physica D: Nonlinear Phenomena} \textbf{237}, 1926--1931.

\bibitem{chae2010lagrangian}
Chae D. 2010  On the Lagrangian dynamics of the axisymmetric 3D Euler
  equations. {\em Journal of Differential Equations} \textbf{249}, 571--577.

\bibitem{Chae_etal_2012}
Chae D, Constantin P, Wu J. 2012  Deformation and Symmetry in the Inviscid
  {SQG} and the {3D} {E}uler Equations. {\em J. Nonlinear Sci.} \textbf{22},
  665--688.

\bibitem{Kiselev_Sverak_2014}
Kiselev A, {\v{S}}ver{\'a}k V. 2014  Small scale creation for solutions of the
  incompressible two-dimensional {E}uler equation. {\em Ann. Math.}
  \textbf{334}, 1205--1220.

\bibitem{CKY_2015}
Choi K, Kiselev A, Yao Y. 2015  Finite Time Blow Up for a {1D} Model of {2D}
  Boussinesq System. {\em Commun. Math. Phys.} \textbf{334}, 1667--1679.

\bibitem{Kiselev_Tan_2018}
Kiselev A, Tan C. 2018  Finite time blow up in the hyperbolic Boussinesq
  system. {\em Adv. Math.} \textbf{325}, 34--55.

\bibitem{Choi_etal_2017}
Choi K, Hou TY, Kiselev A, Luo G, Sverak V, Yao Y. 2017  On the {Finite-Time}
  Blowup of a {One-Dimensional} Model for the {Three-Dimensional} Axisymmetric
  {E}uler Equations. {\em Comm. Pure Appl. Math.} \textbf{70}, 2218--2243.

\bibitem{Bustamante_Kerr}
Bustamante MD, Kerr RM. 2008  3D {E}uler about a 2D symmetry plane. {\em
  Physica D: Nonlinear Phenomena} \textbf{237}, 1912 -- 1920.
Euler Equations: 250 Years On.

\bibitem{Majda_Bertozzi}
Majda AJ, Bertozzi AL. 2002 {\em Vorticity and incompressible flow}.
Cambridge University Press.

\bibitem{Hou_Li_2007}
Hou TY, Li R. 2007  Computing nearly singular solutions using pseudo-spectral
  methods. {\em J. Comput. Phys.} \textbf{226}, 379--397.

\end{thebibliography}

\end{document}